\documentclass{jfm}

\usepackage{graphicx}
\usepackage{newtxtext}
\usepackage{newtxmath}
\usepackage{natbib}
\usepackage{hyperref}
\usepackage{bm}
\hypersetup{
    colorlinks = true,
    urlcolor   = blue,
    citecolor  = black,
}

\newcommand{\RomanNumeralCaps}[1]
\linenumbers

\title{Non-monotonic variations in pressure drop and chaos in viscoelastic fluid flows through an ordered microporous medium}

\author{A. Chauhan\aff{1},
 C. Sasmal\aff{1}
 \corresp{\email{csasmal@iitrpr.ac.in}}}

\affiliation{\aff{1}Department of Chemical Engineering, Indian Institute of Technology Ropar, Rupnagar, Punjab, India-140001}

\begin{document}
\maketitle

\begin{abstract}
Several experimental and numerical studies have shown that viscoelastic fluid flow through a microporous medium exhibits far more complex flow dynamics than Newtonian flow, driven by polymeric elastic stresses that can transition into a chaotic elastic turbulence regime. Recent experiments further revealed that both the chaotic nature of the flow and the pressure drop (\textit{Haward et al., Stagnation points control chaotic fluctuations in viscoelastic porous media flow, PNAS, 2021}; \textit{Browne and Datta, Elastic turbulence generates anomalous flow resistance in porous media, Science Advances, 2021}) across porous media vary non-monotonically with the Weissenberg number (which quantifies the extent of fluid elasticity), although the mechanisms responsible for these observations remain unresolved. The present numerical study aims to investigate and hypothesise about the physical mechanisms governing these complex flow behaviours in an ordered microporous medium consisting of cylindrical micropillars arranged in a staggered configuration. We propose that birefringent strands of high elastic stress, generated by the stretching and alignment of polymer molecules within the porous structure, play the dominant role in controlling the non-monotonic variations in chaos and pressure drop. At low Weissenberg numbers, these stress strands develop gradually, mainly downstream of the micropillars, whereas beyond a critical Weissenberg number, they begin to fluctuate strongly, leading to chaotic flow dynamics. However, at even higher Weissenberg numbers, the strands become larger and stronger, eventually interconnecting between neighbouring micropillars, causing the flow to reorganise into a nearly steady, ordered state similar to that observed at low Weissenberg numbers. On the other hand, the pressure drop in the system consists of a mean contribution, obtained from statistically stationary flow quantities, and a fluctuating contribution. While the mean component increases monotonically with the Weissenberg number, the fluctuating component varies non-monotonically, leading to a similar trend in the total pressure drop. Moreover, the non-monotonic chaotic behaviour strongly depends on the solid volume fraction of the porous medium, which alters both the critical Weissenberg number for instability onset and the range over which the non-monotonic behaviour persists. The findings of the present study could be useful for applications involving viscoelastic fluid flows in porous media, particularly where elastic-turbulence-induced mixing and transport can be exploited to enhance chemical reactions or improve fluid displacement processes.   
\end{abstract}


\section{Introduction}
The topic of flow of viscous fluids through a porous medium can be seen in several applications, such as in oil and water flow in petroleum engineering, groundwater flow and contamination transport in hydrology and environmental science, flow in packed bed reactor in chemical and process engineering, flow of interstitial fluids or drugs in tissues in biomedical engineering, flow in porous electrodes of batteries and fuel cells in material science and energy storage, to name a few~\citep{ingham2005transport,vafai2015handbook}. Due to its numerous applications in various industrial and environmental settings, a substantial body of research has been conducted on this topic over the past several decades. The flow dynamics in a porous medium is significantly complex and rich in physics due to its dependence on several factors, including heterogeneity in micropore network structures, scales of micropores, flow conditions, flow type, i.e., single or multiphase, external fields, etc. Apart from these, fluid properties, particularly non-Newtonian effects such as shear-thinning, shear-thickening, viscoplasticity, and viscoelasticity, were found to significantly affect flow dynamics in porous media. The effect of this particular factor has not just been studied from a fundamental or theoretical point of view, but from the fact that most of the fluids encountered in porous media, in any particular application, often exhibit complex non-Newtonian behaviours. For instance, polymer or surfactant solutions used in enhanced oil recovery (EOR) processes in petroleum engineering, groundwater remediation processes, or filtration processes in chemical and process engineering exhibit viscoelastic and shear-thinning non-Newtonian behaviour. 

The study on the flow of non-Newtonian fluids, particularly viscoelastic fluids, through porous media began much earlier, in the 1960s, when Slattry~\citep{slattery1967flow}, probably the first, modified Darcy's law for predicting the pressure drop of viscoelastic fluids through porous media based on local volume averaging of continuity and momentum equations. Subsequently, Marshall and Metzner~\citep{marshall1967flow} performed experiments with polyisobutylene (PIB) and ET-597 (a variant of high molecular weight polyacrylamide) viscoelastic polymer solutions in a porous medium consisting of a sintered bronze porous disk, and found an increase in pressure drop well above that predicted for purely viscous fluids. Wissler~\citep{wissler1971viscoelastic} showed that this increment in pressure drop can be captured by multiplying a factor $\left(1 + A \left( \frac{\lambda V}{r}\right)\right)^{2}$ to the purely viscous fluid value, where $A$ is a constant to be of order 10, $\lambda$ is the characteristic relaxation time for the fluid, $V$ is the interstitial velocity, and $r$ is a measure of the minimum pore size. An experimental study by Deiber and Schowalter~\citep{deiber1981modeling} in a tube with sinusoidal axial variations in diameter, representing a model porous medium, also found an increase in pressure drop when a polyacrylamide viscoelastic solution flowed through this geometry, consistent with their theoretical analysis. Barboza et al.~\citep{barboza1979viscoelastic} also found an increment in pressure drop in their experiments and proposed that this is attributed to the elongational stresses present in viscoelastic fluids, and accordingly, they suggested the modification of Darcy's law by incorporating an apparent viscosity that will consider the viscosity originated from these elongational stresses. Galindo-Rosales et al.~\citep{galindo2012microfluidic} performed experiments in a model porous medium consisting of a straight microchannel with a sequence of contractions and expansions. They also observed a substantial increase in the pressure drop in this model porous system for the flow of viscoelastic polymer solutions, as seen in their experiments with an actual porous medium consisting of a cylindrical tube filled with sand. Recently, Carlson et al.~\citep{carlson2022volumetric} conducted an experimental study in a microscale three-dimensional ordered porous medium and performed a detailed fluid flow visualization and analysis at the pore level with micro-tomographic particle image velocimetry $(\mu-\text{TPIV})$. They also found a substantial increase in the pressure drop once the Weissenberg number (which quantifies the extent of fluid viscoelasticity) exceeds a critical value of 2, due to the onset of the chaotic elastic turbulence (ET) regime, as they explained.  Likewise experiments, De et al.~\citep{de2017viscoelastic} also found a similar trend in their numerical simulations for a three-dimensional random porous medium consisting of spheres packed at different volume fractions, wherein they reported this steep increase in terms of an apparent viscosity $\eta_{\text{app}}$, defined as $\frac{\left( \frac{\Delta P}{<u>}\right)_{\text{VE}}}{\left( \frac{\Delta P}{<u>}\right)_{\text{N}}}$, where $\Delta P$ is the pressure drop, $<u>$ is the volume-averaged velocity, and $\text{VE}$ and $\text{N}$ stands for viscoelastic and Newtonian fluids, respectively. Although all prior studies found an increase in the pressure drop in viscoelastic fluids compared to that in simple Newtonian viscous fluids, interestingly, Browne and Datta~\citep{browne2021elastic} found a more complex dependence between the pressure drop and the Weissenberg number in their experiments with a three-dimensional porous medium consisting of a squared-section quartz capillary filled with borosilicate glass beads. Initially, they also observed an anomalous increase in the pressure drop as the Weissenberg number increased gradually, as seen in prior studies. It reached a maximum value at a critical Weissenberg number, and then started to decrease as the Weissenberg number further increased, resulting in a non-monotonic trend between the two instead of a monotonic one.          

On the other hand, apart from a complex pressure drop variation, the flow of viscoelastic fluids often causes the generation of the elastic instability, driven by elastic stresses, once the Weissenberg number exceeds a critical value~\citep{larson1990purely,pakdel1996elastic}. This instability can transit to a more chaotic state, the so-called elastic turbulence (ET) regime, on further increasing the Weissenberg number, resulting in the origin of a chaotic flow field, reminiscent of the inertial turbulence, inside the porous medium, even in the low Reynolds number regime~\citep{datta2022perspectives,steinberg2021elastic,groisman2000elastic,sasmal2025potential}. Recently, a series of studies have been conducted to investigate how the geometric parameters associated with a porous medium, such as micropore shape and separation distance, and the arrangement of micropillars often used to mimic a porous medium, influence the transition of the flow field from a stable and ordered state to an unstable and chaotic state during the flow of viscoelastic fluids. For instance, Browne et al.~\citep{browne2020bistability} performed an experimental study with a one-dimensional ordered array of micropore constrictions for representing a porous medium, and found the presence of multistability, i.e., stochastically switching among distinct unstable flow states when the distance between two consecutive pores is small. Subsequently, a recent study by Chen and Datta~\citep{chen2025influence} from the same research group investigated how fluid rheological properties, namely, shear-thinning and elasticity, influence this multistability, and found that the fluid must be sufficiently elastic for the onset of this flow transition. Galindo-Rosales et al.~\citep{galindo2012microfluidic} found the existence of different vortices, which they categorised as small, growing, and asymmetric vortices, inside the pores before the flow transits to a more unstable state. Further studies were also conducted to investigate the flow dynamics of viscoelastic fluids, such as polymer or surfactant solutions, in a single pore, and also showed the existence of several flow states and vortical structures inside it~\citep{kumar2021numerical,boek2007flow,browne2020pore,sasmal2020flow}.  

Therefore, it can be seen that a flow transition from a stable to an unstable state occurs in viscoelastic porous media flows after a critical value is reached, either in the flow rate or in the fluid's rheological properties, particularly its elasticity. This ultimately leads to the generation of chaos inside the porous medium. Walkama et al.~\citep{walkama2020disorder} performed an experimental study and found that introducing geometric disorder in a model porous medium, consisting of cylindrical micropillars in a straight microchannel arranged in a staggered manner, suppresses this chaotic flow behaviour. They proposed that disorder introduces preferential flow paths that promote shear over extensional deformation and enhance flow stability by locally reducing polymer stretching. Subsequently, Haward et al.~\citep{haward2021stagnation} performed an experimental study on the same model porous medium but with micropillars initially arranged in an aligned manner. They observed the origin of chaotic flow behaviour in this flow setup upon introducing geometric disorder, in contrast to that reported by Walkama et al. Based on these two contrasting results, they concluded that the chaotic flow behaviour in a porous medium is controlled by the number of stagnation points on the micropillars exposed to the flow. Following these experimental studies, further numerical studies have been conducted to unravel the mechanisms underlying the onset and suppression of this chaotic flow behaviour. For instance, in two back-to-back studies, Mokhtari et al.~\citep{mokhtari2024web,mokhtari2022birefringent} demonstrated that the flow in a model porous medium is guided by the birefringent strands, and the localised stress accumulated due to these strands controls the spatio-temporal flow fluctuations. Chauhan et al.~\citep{chauhan2022effect} suggested that the formation of preferential paths and the stretching of polymers in those paths lead to the onset of these chaotic fluctuations, which are again greatly influenced by the shear-thinning properties of the fluid.

One important aspect is whether these chaotic fluctuations continue to grow as the flow rate or fluid elasticity increases. This is because Haward et al.~\citep{haward2021stagnation} find that there is a non-monotonic trend present in these chaotic fluctuations with the Weissenberg number in their experiments. They first grow, reach a maximum, and then decrease, even in a regular staggered arrangement of micropillars used to create the porous medium. Likewise, chaotic fluctuations, as mentioned earlier, Browne and Datta~\citep{browne2021elastic} also found a non-monotonic trend in the pressure drop, even in a three-dimensional porous medium composed of randomly packed spheres. Therefore, one of the key questions is: What is the mechanism behind this non-monotonic behaviour, either in chaotic flow fluctuations or in pressure drop variations in viscoelastic porous media flows? This study aims to address this important question through extensive numerical simulations and to identify a possible underlying mechanism not yet reported in the literature by considering an ordered porous medium consisting of cylindrical micropillars arranged in a staggered manner. This is highly relevant to investigate, enabling one to better control chaotic fluctuations or pressure drop variations in this particular flow problem from an application standpoint, associated with numerous industrial, environmental, and biological settings.

\section{Problem description}\label{ProbDes}

\begin{figure}
    \centering
    \includegraphics[width=14cm]{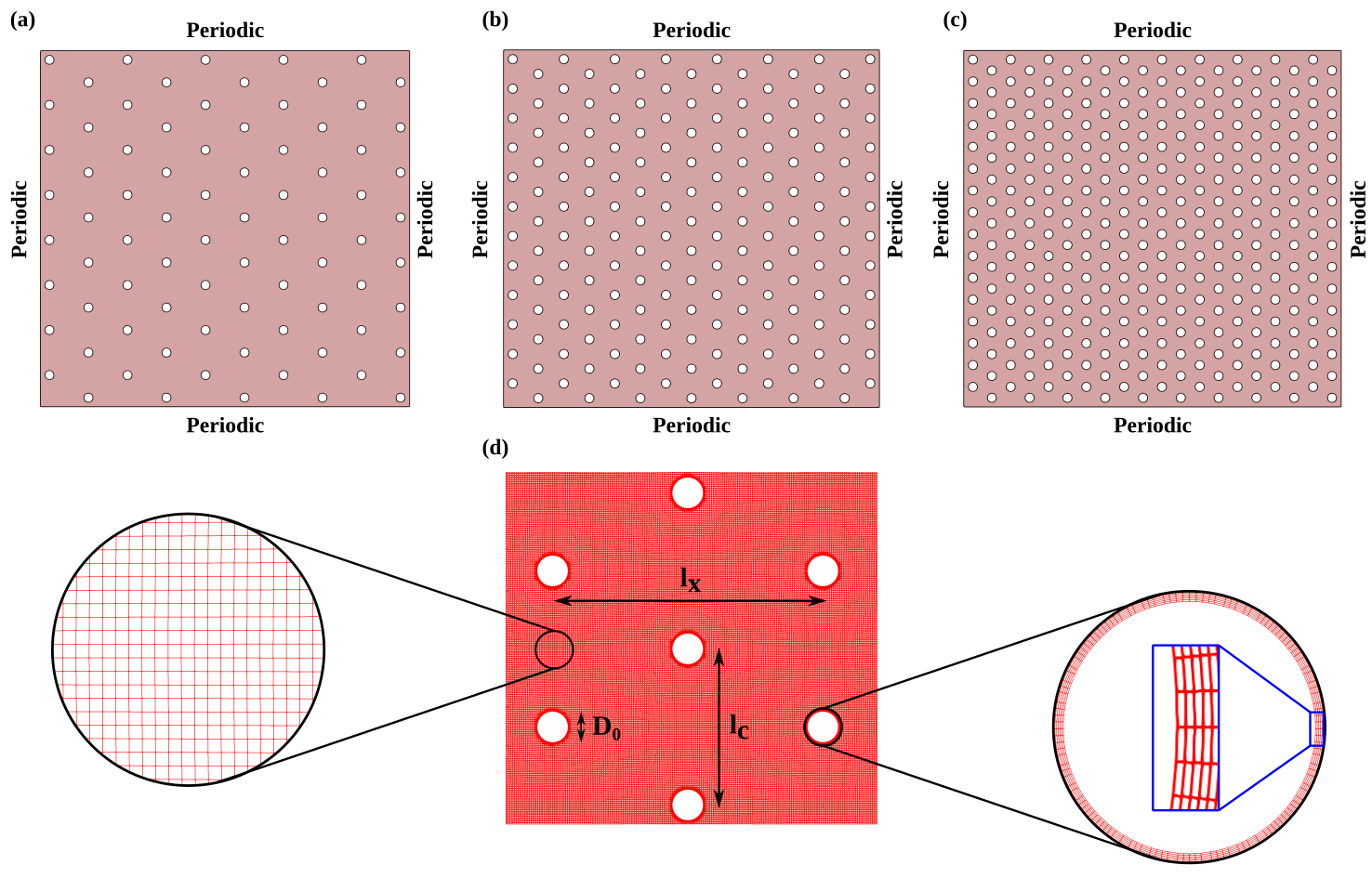}
    \caption{Schematic of the periodic porous domain considered in the present study for three different solid volume fractions $(\phi)$: $\phi \approx 3.88 \%$ with 80 pillars (a), $\phi \approx 8.93 \%$ with 180 pillars (b), and $\phi \approx 15.45 \%$ with 320 pillars (c). Sub-figure (d) shows the typical grid structure for one hexagonal array of micropillars with an appropriate zoom-in on either side. For the ease of understanding, labelling is also done here for the micropillar diameter as `$D_0$', streamwise spacing as `$l_x$', and centre-to-centre spacing as `$l_c$', whose values are provided in table~\ref{table:grid_details}.}
    \label{schematicAndGrid}
\end{figure}

As mentioned in the preceding section, this study aims to investigate the flow of viscoelastic fluids through an ordered microporous medium consisting of cylindrical micropillars arranged in a staggered manner, as schematically shown figure~\ref{schematicAndGrid}. Three different solid volume fractions of micropillars within the porous medium are considered in this study: $3.88\%$ with 80 micropillars, $8.93\%$ with 180 micropillars, and $15.45\%$ with 320 micropillars, as shown in the same figure. Periodic boundary conditions are applied at the edges of the computational domain to simulate a large medium in which the flow is driven from left to right with a velocity $u_0$. Furthermore, the flow is assumed to be incompressible in nature, and under this condition, it is governed by the continuity and momentum equations written in their dimensional forms as below:
\newline
Continuity equation:
\begin{equation}\label{eq:continuity_dimensional}
\bm{\nabla^*}\cdot\bm{u^*}=0 
\end{equation}
Cauchy's momentum equation:
\begin{equation}\label{eq:momentum_dimensional}
\rho \left(\dfrac{\partial{\bm{u^*}}}{\partial{t^*}}+ \bm{u^*}\cdot\bm{\nabla^*}\bm{u^*}\right) = -\bm{\nabla^*}{p^*} + \bm{\nabla^*}\cdot\bm{\tau^*_{s}}+\bm{\nabla^*}\cdot\bm{\tau^*_{p}}
\end{equation}
Here, $\bm{\nabla^*}$ is the gradient operator, $\bm{u^*}$ is the velocity vector, $\rho$ is the fluid density, $t^*$ is time, $p^*$ is pressure, $\bm{\tau^*_{s}}$ is the solvent contribution to the stress tensor, and $\bm{\tau^*_{p}}$ is the polymeric contribution to the stress tensor. In the above equations, $<*>$ in the superscript denotes a dimensional variable. Moreover, the FENE-CR (finitely extensible non-linear elastic spring-Chilcott and Rallison) viscoelastic model~\citep{chilcott1988creeping} is the constitutive model chosen in the present study to represent the polymeric contribution appearing in the last term of the momentum equation. This viscoelastic model, in particular, is suitable for investigating the effect of fluid elasticity alone on flow dynamics, which is the main aim of the present study. Not only this, but the FENE-CR model also accounts for the finite extensibility of the polymer molecule, unlike the other viscoelastic models, such as the Oldroyd-B model~\citep{shaqfeh2021oldroyd}, which allows for the infinite stretching of polymer molecules, further making the FENE-CR model more realistic from a practical point of view. Furthermore, the mathematical expressions to evaluate $\bm{\tau^*_s}$ and $\bm{\tau^*_p}$ are expressed as: 

\begin{equation}\label{eq:solvent_stress_dimensional}
    \bm{\tau^*_s} = 2 \eta_s \bm{\dot{\gamma}^*}
\end{equation}

\begin{equation}\label{eq:polymeric_transport_dimensional}
\left[1+\lambda \frac{D}{Dt^*}\left(\frac{1}{f^*}\right)\right]\bm{\tau^*_p} + \frac{\lambda}{f^*}\overset{\bm{\nabla}}{\bm{\tau^*_p}} = \eta_p \left[\bm{\nabla^*} \bm{u^*} + (\bm{\nabla^*} \bm{u^*})^T \right]
\end{equation}
Where, $\eta_s$ is the solvent viscosity, $\bm{\dot{\gamma}^*} = \frac{1}{2} \left(\bm{\nabla^*} \bm{u^*} + (\bm{\nabla^*} \bm{u^*})^T \right)$ is the deformation-rate tensor, $\lambda$ is the relaxation time of the polymer molecule, $\frac{D (.)}{Dt^*} = \frac{\partial (.)}{\partial t^*} + \bm{u^*}\cdot\bm{\nabla^*} (.)$ is the material derivative of any variable $(.)$, $f^* = \frac{L^2 + \frac{\lambda}{\eta_p} tr(\bm{\tau^*_p})}{L^2 - 3}$, $\overset{\bm{\nabla}}{\bm{\tau^*}_{p}} = \dfrac{\partial{\bm{\tau^*}_{p}}}{\partial{t^*}}+\bm{u^*}\cdot\bm{\nabla^*}\bm{\tau^*}_{p}-\bm{\tau^*}_{p}\cdot\bm{\nabla^*}\bm{u^*}-(\,\bm{\bm{\nabla^*}}\bm{u^*})^T\,\cdot\bm{\tau^*}_{p}$ is the upper-convected derivative of $\bm{\tau}^*_p$, $\eta_p$ is the polymeric viscosity, and $L^2$ is the polymer extensibility parameter.

The various terms involved in the above set of equations (equations~\ref{eq:continuity_dimensional},~\ref{eq:momentum_dimensional},~\ref{eq:solvent_stress_dimensional} and \ref{eq:polymeric_transport_dimensional}) are non-dimensionalised using the following scaling variables: length is scaled with micropillar diameter $D_0$, velocity is scaled with $u_0$, time is scaled with $D_0/u_0$, pressure, solvent stresses and polymeric stresses with $\eta_0 u_0/D_0$. Here, $\eta_0$ is the zero-shear rate viscosity of the resulting viscoelastic solution, given as $\eta_0 = \eta_s + \eta_p$ with $\eta_s$ and $\eta_p$ are the solvent and polymeric constributions, respectively. The corresponding non-dimensional forms of the above-stated governing equations become as follows: 

\begin{equation}\label{eq:continuity_non-dimensional}
\bm{\nabla}\cdot\bm{u}=0 
\end{equation}

\begin{equation}\label{eq:momentum_non-dimensional}
Re \left(\dfrac{\partial{\bm{u}}}{\partial{t}}+ \bm{u}\cdot\bm{\nabla}\bm{u}\right) = -\bm{\nabla}{p} + \bm{\nabla}\cdot\bm{\tau_{s}}+\bm{\nabla}\cdot\bm{\tau_{p}}
\end{equation}

\begin{equation}\label{eq:solvent_stress_non-dimensional}
    \bm{\tau_s} = 2 \beta \bm{\dot{\gamma}}
\end{equation}

\begin{equation}\label{eq:polymeric_transport_non-dimensional}
\left[1+ Wi \frac{D}{Dt}\left(\frac{1}{f}\right)\right]\bm{\tau_p} + \frac{Wi}{f}\overset{\bm{\nabla}}{\bm{\tau_p}} = (1-\beta) \left[\bm{\nabla} \bm{u} + (\bm{\nabla} \bm{u})^T \right]
\end{equation}
Where, $Re = \frac{\rho u_0 D_0}{\eta_0}$ is the Reynolds number defined as the ratio of inertial forces to the viscous forces, $Wi = \frac{\lambda u_0}{D_0}$ is the Weissenberg number defined as the ratio of the polymer relaxation time to the time scale of the flow, $\beta = \frac{\eta_s}{\eta_0}$ is the viscosity ratio, and $f = \frac{L^2 + \frac{Wi}{1-\beta} \text{tr}(\bm{\tau_p})}{L^2 - 3}$. Note that, for a Newtonian fluid, $\beta \rightarrow 1$. Furthermore, in the present study, we explicitly solve for the polymeric conformation tensor $\bm{C}$ rather than $\bm{\tau_p}$ to account for the viscoelastic fluid's flow deformation history. For the FENE-CR model, $\bm{\tau_p} = \left(\frac{1-\beta}{Wi}\right) \left(\frac{L^2}{L^2 - tr(\bm{C})}\right) \left(\bm{C}-\bm{I}\right)$. On introducing $\bm{C}$ in the polymeric transport equation (equation~\ref{eq:polymeric_transport_non-dimensional}), the constitutive relation reduces to $\bm{C} + \frac{Wi}{L^2/(L^2-tr(\bm{C}))}\overset{\bm{\nabla}}{\bm{C}} = \bm{I}$. However, it is worth noting that the numerical solution of this equation often diverges beyond a critical value of the Weissenberg number, primarily due to the formation of high or steep gradients and the loss of positive definiteness of the stress tensor. Such numerical divergence often arises in complex geometries with geometric singularities, as in the present case of porous media flows. This problem is popularly known as the High-Weissenberg Number Problem (HWNP)~\citep{keunings1986high}. As a remedy, Fattal and Kupferman~\citep{fattal2005time} introduced the log-conformation tensor approach, which was later implemented in OpenFOAM by Pimenta and Alves~\citep{pimenta2017stabilization}. As per the procedure, a new tensor ($\bm{\Theta}$) is introduced, which is defined as the natural logarithm of the conformation tensor $\bm{C}$, i.e., $\bm{\Theta} = \text{ln}(\bm{C}) = \bm{B}~\text{ln}(\bm{A}) \bm{B}^T$. Owing to the fact that $\bm{C}$ is a positive-definite tensor, it is diagonalized as $\bm{C} = \bm{B}\bm{A}\bm{B}^T$, where $\bm{B}$ is a matrix that consists the eigenvectors of $\bm{C}$ in its columns and the corresponding eigenvalues resulting from the decomposition of $\bm{C}$ are contained in the diagonal elements of matrix $\bm{A}$. The transformation of $\bm{C}$ to $\bm{\Theta}$ ultimately results in the following constitutive equation for the FENE-CR model:

\begin{equation}
\label{eq:log-conformation_transport}
\dfrac{\partial{\bm{\Theta}}}{\partial{t}}+\bm{u}\cdot\bm{\nabla}\bm{\Theta}=\bm{\Omega}\bm{\Theta}-\bm{\Theta}\bm{\Omega}+2\bm{D}+\frac{L^2/(L^2-tr(\bm{\bm{\Theta}}))}{Wi}(e^{-\bm{\Theta}}-\bm{I})
\end{equation}
where,
\begin{equation}
\bm{D} = \bm{B} \begin{bmatrix} m_{xx} & 0 & 0 \\ 0 & m_{yy} & 0 \\ 0 & 0 & m_{zz}  \end{bmatrix} \bm{B}^T
\end{equation}

\begin{equation}
\bm{\Omega} = \bm{B} \begin{bmatrix} 0 & \Psi_{xy} & \Psi_{xz} \\ -\Psi_{xy} & 0 & \Psi_{yz} \\ -\Psi_{xz} & -\Psi_{yz} & 0 \end{bmatrix} \bm{B}^T
\end{equation}

\begin{equation}
\bm{M} = \bm{B}^{T} (\bm{\nabla} \bm{u})^{T} \bm{B} = \begin{bmatrix} m_{xx} & m_{xy} & m_{xz} \\ m_{yx} & m_{yy} & m_{yz} \\ m_{zx} & m_{zy} & m_{zz} \end{bmatrix}
\end{equation}

\begin{equation}
\Psi_{ij} = \frac{A_{j} m_{ij} + A_{i} m_{ji}}{A_j - A_i}
\end{equation}
The solution of equation~\ref{eq:log-conformation_transport} is obtained in the diagonalized form as $\bm{\Theta} = \bm{B} \bm{A}^{\bm{\Theta}} \bm{B}^{T}$ and the recovery of $\bm{C}$ takes place with the help of inverse relation as $\bm{C} = e^{\bm{\Theta}} = \bm{B} e^{\bm{A}^{\bm{\Theta}}} \bm{B}^{T}$. Later on, the correlation $\bm{\tau_p} = \left(\frac{1-\beta}{Wi}\right) \left(\frac{L^2}{L^2 - tr(\bm{C})}\right) \left(\bm{C}-\bm{I}\right)$ is utilized to estimate $\bm{\tau_p}$, which is ultimately used in the momentum equation (equation~\ref{eq:momentum_non-dimensional}).

\section{Numerical details}\label{Numerical_details}

To numerically solve the governing equations outlined in the preceding section, including the continuity and momentum equations, we use the finite volume method (FVM) in the open-source computational fluid dynamics (CFD) toolbox OpenFOAM~\citep{weller1998tensorial}. In addition, we employ the RheoTool package~\citep{rheoTool} to solve the FENE-CR constitutive equations that account for the polymeric stresses in the momentum equation. Regarding the discretisation of various terms appearing in the governing equations, we use the CUBISTA (Convergent and Universally Bounded Interpolation Scheme for Treatment of Advection) scheme~\citep{alves2003convergent} for advective terms appearing in the FENE-CR constitutive equations, whereas we adopt the Gauss linear orthogonal interpolation scheme for handling the diffusive terms, ensuring both schemes are second-order accurate in space. It is worth noting here that in the present study, we have performed simulations for the perfectly creeping flow condition $(Re = 0)$, due to which we have disabled the discretisation of the convection term appearing in the momentum equation. Moreover, the gradient terms and time derivatives were discretised using Gauss linear interpolation and the Euler scheme, respectively. Upon discretisation, the obtained set of linear systems linked with pressure and velocity was solved using PCG (Preconditioned Conjugate Gradient) solver with a DIC (Diagonal-based Incomplete Cholesky) preconditioner, whereas those associated with the stress fields were solved using the PBiCG (Preconditioned Bi-Conjugate Gradient) solver with a DILU (Diagonal-based Incomplete Lower Upper) preconditioner~\citep{lee2003incomplete,ajiz1984robust}. The pressure-velocity coupling was effectuated through the SIMPLE (Semi-Implicit Method for Pressure Linked Equations) algorithm. As mentioned earlier, the simulations in the present study were stabilised using the log-conformation approach~\citep{pimenta2017stabilization}, and to ensure robustness of the solver, a relative tolerance of $10^{-10}$ was set for all field variables, such as pressure, velocity, and stresses.

\begin{table}
  \caption{Details related to the various grid structures adopted in the present study for three different solid volume fraction values.}
  \label{table:grid_details}
  \begin{center}
  \scalebox{0.65}{
  \begin{tabular}{cccccccccc}
    \hline
    {} & \multicolumn{3}{c}{$\phi \approx 3.88\%$} & \multicolumn{3}{c}{$\phi \approx 8.93\%$} & \multicolumn{3}{c}{$\phi \approx 15.45\%$} \\
    {} & Grid-1 & Grid-2 & Grid-3 & Grid-1 & Grid-2 & Grid-3 & Grid-1 & Grid-2 & Grid-3 \\
    \hline
    Total number of micropillars & \multicolumn{3}{c}{80} & \multicolumn{3}{c}{180} & \multicolumn{3}{c}{320} \\
    Streamwise micropillar spacing, $l_x/D_0$ & \multicolumn{3}{c}{$8.66$} & \multicolumn{3}{c}{$5.54$} & \multicolumn{3}{c}{$4.16$} \\
    Centre-to-centre spacing, $l_c/D_0$ & \multicolumn{3}{c}{$5.00$} & \multicolumn{3}{c}{$3.20$} & \multicolumn{3}{c}{$2.40$} \\
    Cell thickness near micropillar, $\Delta r/D_0$ & $6.35 \times 10^{-3}$ & $5.58 \times 10^{-3}$ & $4.63 \times 10^{-3}$ & $7.93 \times 10^{-3}$ & $5.57 \times 10^{-3}$ & $4.56 \times 10^{-3}$ & $6.47 \times 10^{-3}$ & $5.63 \times 10^{-3}$ & $4.68 \times 10^{-3}$ \\
    Minimum grid cell area, $A_{\text{min}}/D^2_0$ & $1.20 \times 10^{-4}$ & $8.96 \times 10^{-5}$ & $6.15 \times 10^{-5}$ & $1.73 \times 10^{-4}$ & $8.69 \times 10^{-5}$ & $6.00 \times 10^{-5}$ & $1.20 \times 10^{-4}$ & $8.91 \times 10^{-5}$ & $6.12 \times 10^{-5}$ \\
    Maximum grid cell area, $A_{\text{max}}/D^2_0$ & $4.13 \times 10^{-3}$ & $3.09 \times 10^{-3}$ & $2.14 \times 10^{-3}$ & $5.73 \times 10^{-3}$ & $2.88 \times 10^{-3}$ & $1.99 \times 10^{-3}$ & $3.80 \times 10^{-3}$ & $2.85 \times 10^{-3}$ & $1.98 \times 10^{-3}$ \\    
    Total number of grid cells & 612,352 & 810,729 & 1,151,303 & 464,613 & 861,354 & 1,202,252 & 728,626 & 932,184 & 1,273,906 \\
    Maximum mesh non-orthogonality & 55.65 & 56.15 & 55.82 & 56.75 & 56.22 & 55.68 & 57.21 & 56.46 & 56.09 \\
    Average mesh non-orthogonality & 3.13 & 2.85 & 2.59 & 5.02 & 4.20 & 3.82 & 5.76 & 5.38 & 4.95 \\
    Maximum mesh skewness & 0.55 & 0.56 & 0.55 & 0.57 & 0.55 & 0.55 & 0.57 & 0.57 & 0.56 \\
    \hline
  \end{tabular}}
  \end{center}
\end{table}

\begin{figure}
    \centering
    \includegraphics[width=10cm]{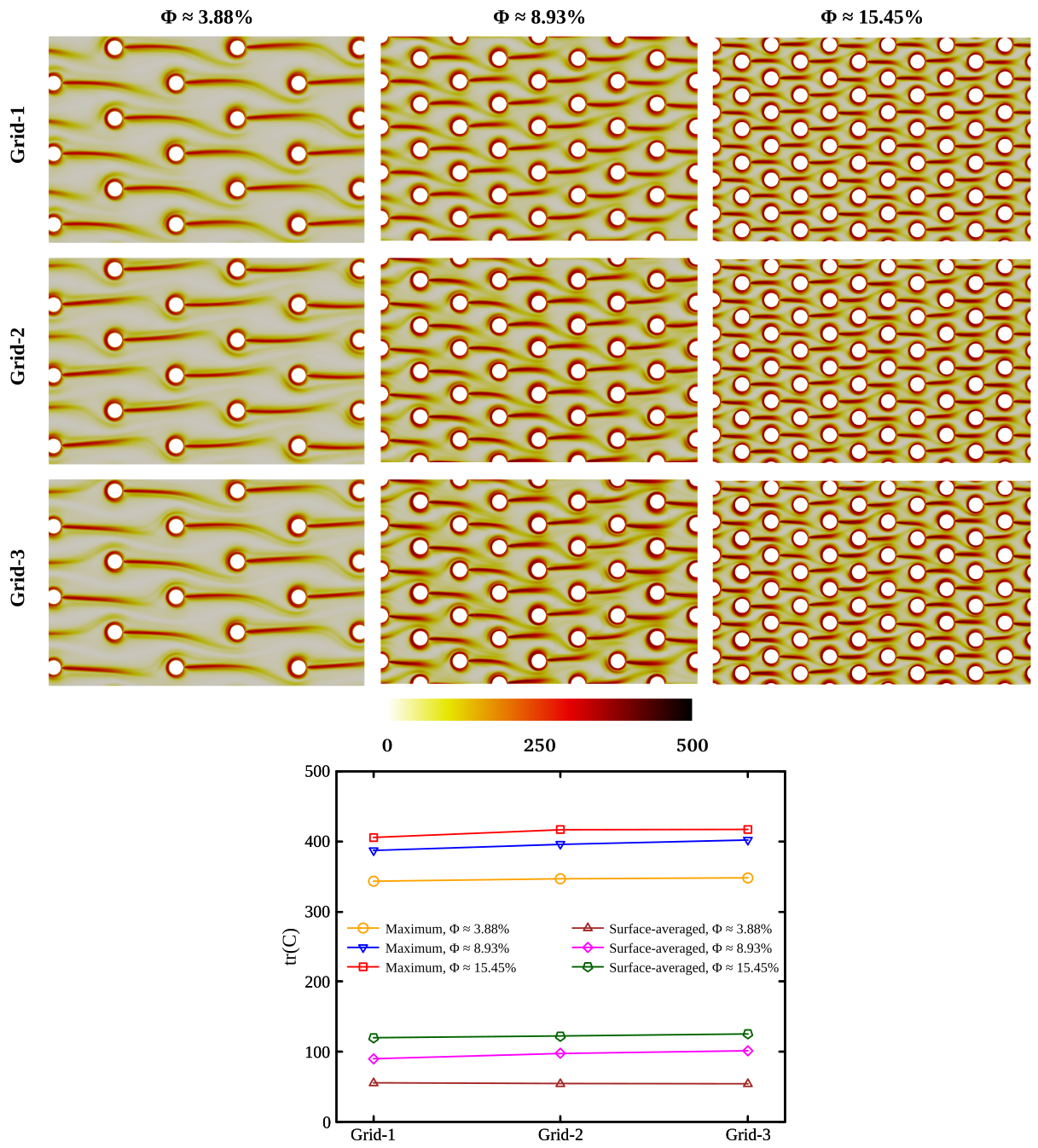}
    \caption{Contours of $tr(\bm{C})$ at three different grid densities and solid volume fraction values used in this study. The sub-figure embedded at the bottom denotes the maximum and surface-averaged values of $tr(\bm{C})$ for the domain shown above. Here, the parameter values for the FENE-CR model are $Wi = 5$, $\beta = 0.75$, and $L^2 = 500$.}
    \label{gridTest}
\end{figure}

In addition to scrutinising various discretisation techniques and tolerance levels, careful attention was paid while selecting the grid density and time-step size in the present simulations. Specifically, the background grid was first generated for the computational domain using the \textit{blockMeshDict} subroutine in OpenFOAM. Later on, the \textit{snappyHexMesh} utility was adopted to place multiple circular micropillars, thereby constructing the resulting porous media as shown in sub-figure~\ref{schematicAndGrid}(d). It should be noted that a relatively fine grid was constructed in the vicinity of all circular micropillars to capture the steep gradients of velocity, pressure, stress, etc. in the present study. In particular, 5 circular layers were added close to the surface of the micropillars as shown in the zoom-in of the sub-figure~\ref{schematicAndGrid}(d) on the right-hand side. Detailed information on three different grid structures, constructed for each solid volume fraction, is presented in the table~\ref{table:grid_details}. The effect of grid density is portrayed in figure~\ref{gridTest}, wherein the contours of $tr(\bm{C})$ are calculated for all three grids and solid volume fraction values. This parameter is selected to determine the current grid because the central conclusion of this study is based on the dynamics of the birefringent strands. Evidently, the fluctuating birefringent strands formed downstream of the micropillars look similar to one another for each solid volume fraction value (see each column for grid variation in figure~\ref{gridTest}). To assess it quantitatively, we calculate the maximum and surface-averaged values of $tr(\bm{C})$ for the shown domain. This has been included in the same figure~\ref{gridTest} on the bottom side, wherein the variation in both maximum and surface-averaged values of $tr(\bm{C})$ is minimal, particularly when one moves from grid-2 to grid-3. Therefore, the grid-2, with approximately $0.8-0.9$ million cells, is found to be adequate for the present work. Furthermore, a non-dimensional time-step size of $\Delta t = 10^{-3}$ was selected for all simulations performed in the present study. This corresponding dimensional value of the time-step size is $10^{-4}\,s$, which is almost 3000 times smaller than the polymer relaxation time used in this study. Therefore, it is sufficient to accurately capture the flow dynamics. Furthermore, this value of $\Delta t$ is chosen in accordance with the criteria of CFL
(Courant-Friedrichs-Lewy)~\citep{courant1967partial}, typically by monitoring the Courant number $(Co)$, which also ensures the numerical stability. Mathematically, $Co = \frac{u^* \Delta t^*}{\Delta x^*}$, where $\Delta t^*$ is the time-step size used in the present simulations, and $\frac{\Delta x^*}{u^*}$ represents the characteristic convective time-scale of flow. For the numerical solutions to be convergent, $Co \leq 1$. In all simulations conducted in the present study, the maximum value of $Co$ was observed to be 0.1 or less. Finally, the following set of boundary conditions has been imposed: the left and right, and top and bottom sides of the domain are set to be periodic. At the surface of the circular micropillars in the porous domain, the standard no-slip and no-penetration condition for velocity, zero-gradient for pressure, and linear extrapolation for polymeric stress tensor are employed.

\section{Results and discussion}\label{Results}

\subsection{Validation}

\begin{figure}
    \centering
    \includegraphics[width=13cm]{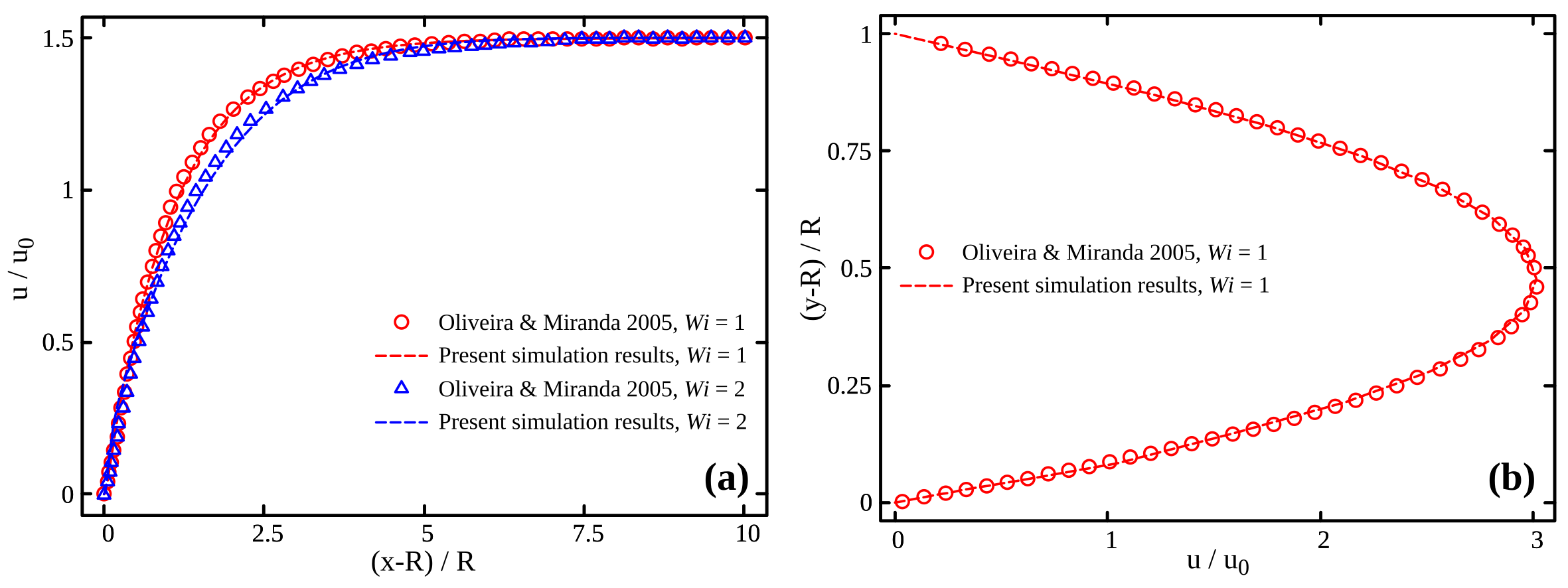}
    \caption{Validation of the present implemented FENE-CR model with the existing numerical results of Oliveira $\&$ Miranda~\citep{oliveira2005numerical} for flow past a bounded cylinder with a blockage ratio of 2. Normalised velocity profile along the mid-horizontal line downstream of the cylinder (a), and along a vertical line passing through the centre of the cylinder (b). Here, the origin is kept at the centre of the cylinder with radius `R'.}
    \label{validation1}
\end{figure}

\begin{figure}
    \centering
    \includegraphics[width=13cm]{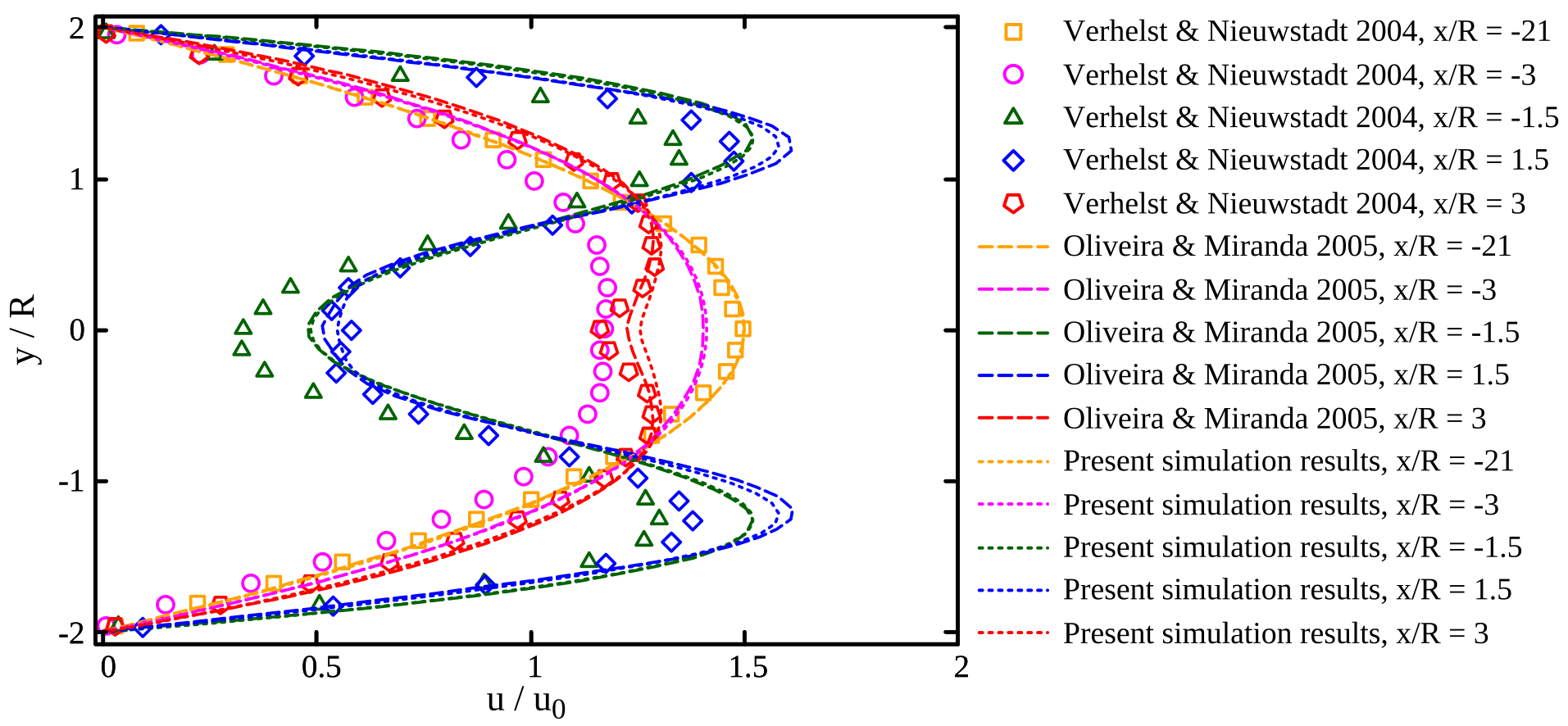}
    \caption{Validation of the present implemented FENE-CR model with the existing experimental results of Verhelst $\&$ Nieuwstadt~\citep{verhelst2004visco} and numerical results of Oliveira $\&$ Miranda~\citep{oliveira2005numerical} for flow past a cylinder confined in a rectangular channel. Here, the results are compared in terms of velocity (normalised by the mean velocity $u_0$) at five different vertical lines such that $\text{x/R} = -21, -3, -1.5, 1.5, \text{and}~3$, where the origin is kept at the centre of the cylinder with radius R = 5 mm.}
    \label{validation2}
\end{figure}

Before presenting the new results, it is imperative to validate the present numerical solver for the FENE-CR viscoelastic model used in this study. To do so, we first carry out a validation study with the numerical results of Oliveira $\&$ Miranda~\citep{oliveira2005numerical}, wherein they used a system of flow past a cylinder confined in a channel with a blockage ratio of 2 in the low Reynolds number regime. The comparison between the two results is shown in figure~\ref{validation1}, and clearly, the results show an excellent agreement for both the streamwise velocity variation along the mid-horizontal line downstream of the cylinder and along a vertical line passing through the centre of the cylinder. In their same paper, Oliveira $\&$ Miranda~\citep{oliveira2005numerical} also presented some comparison against the experimental results of Verhelst $\&$ Nieuwstadt~\citep{verhelst2004visco}, wherein the flow past a cylinder (diameter = 10 mm) confined in a rectangular channel (length = 1100 mm, width = 20 mm, and height = 160 mm) was investigated for a elastic solution without shear-thinning (Boger fluid) effect, prepared using $93\%$ glucose syrup, $7\%$ distilled water and 150 ppm of partially hydrolysed polyacrylamide (PAMH) polymer. Figure~\ref{validation2} shows the comparison of the present numerical results obtained using the FENE-CR model ($Wi = \frac{\lambda u_0}{R} = 1.2, L^2 = 144$ where $u_0$ is the channel inlet velocity and $R$ is the cylinder radius) with the experimental results of Verhelst $\&$ Nieuwstadt~\citep{verhelst2004visco} and numerical results of Oliveira $\&$ Miranda~\citep{oliveira2005numerical}. While the numerical results are again in good agreement, the agreement with the experimental results is also reasonable. These validation studies, along with those presented in our previous study~\citep{chauhan2022effect}, highlight the robustness of the present numerical solver and provide confidence in discussing the new results for viscoelastic porous media flows presented in the ensuing section.

\subsection{Flow field transition at a fixed solid volume fraction}

\begin{figure}
    \centering
    \includegraphics[width=14cm]{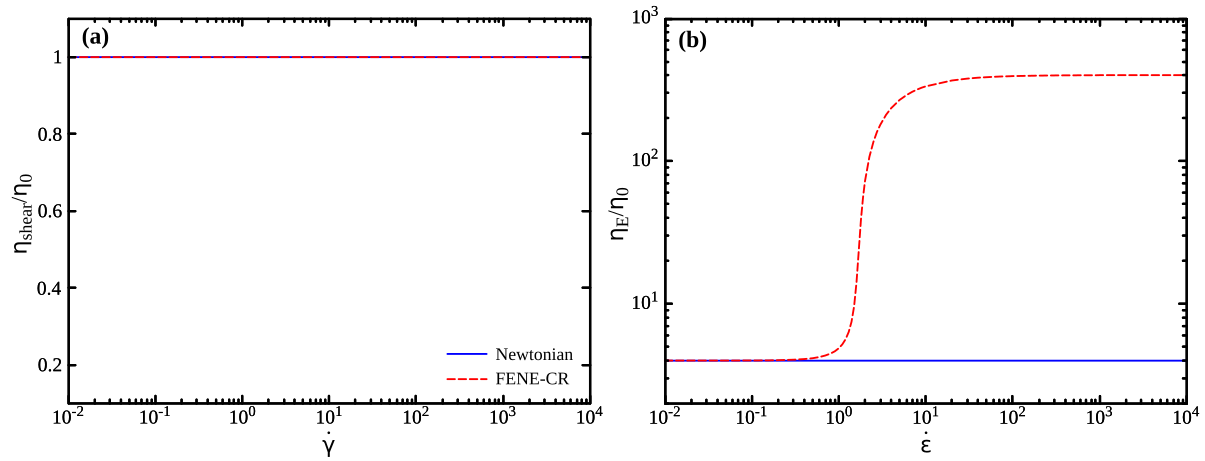}
    \caption{Variation of shear viscosity under steady simple shear flows (a), and extensional viscosity under planar extensional flows (b) for both Newtonian and FENE-CR models.}
    \label{shearAndExtensional}
\end{figure}

To investigate the effect of fluid elasticity on flow transition inside the present flow system at a fixed solid volume fraction, we have varied the Weissenberg number by varying the mean velocity $u_0$ imposed on the flow domain to drive the flow. The values of other dimensional parameters are kept constant as follows: $\lambda = 0.3 s$, $\beta = 0.6$, $L^2 = 500$, and $\phi \approx 8.93 \%$. These selected viscosity ratio and relaxation time values are identical to those reported by McKinley et al.~\citep{mckinley1993wake}. However, we have used a larger value of $L^2 = 500$ to obtain a stronger extensional thickening effect, responsible for elastic instability. A similar range of $L^2$ values has also been used in other studies for the flow past a circular cylinder~\citep{richter2010simulations,oliveira2001method}. The rheological response of the viscoelastic fluid with these parameter values of the FENE-CR constitutive model in simple steady shear and planar extensional flows is shown in sub-Figures~\ref{shearAndExtensional} (a) and (b), respectively. It can be seen that the present viscoelastic fluid exhibits a constant shear viscosity in simple shear flows, whereas an extensional thickening followed by a plateau-like value is observed in planar extensional flows. A Newtonian case (with $Wi = 0$) is also simulated to make a comparison with viscoelastic fluids under otherwise identical conditions. Note that the simulations are carried out for perfectly creeping flows (i.e., $Re = 0$) by disabling the discretisation of the convective terms of the governing equations as outlined in the preceding section. 

\begin{figure}
    \centering
    \includegraphics[width=10cm]{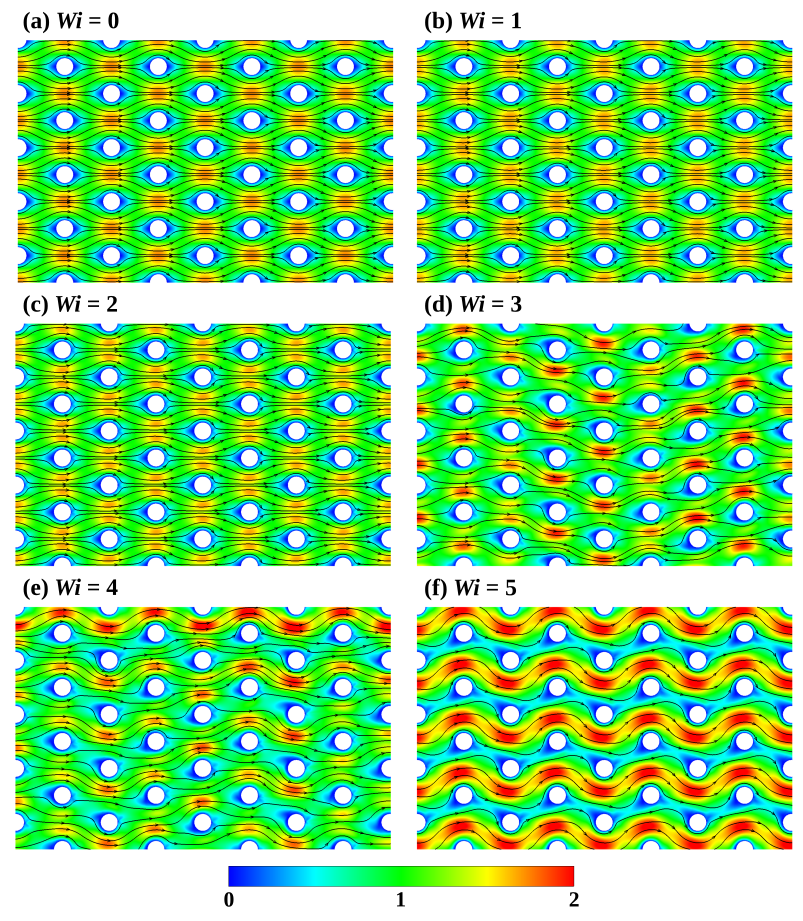}
    \caption{Streamlines and velocity magnitude (normalised by $u_0$) plots at different Weissenberg numbers. Here, the Weissenberg number is changed by varying the mean velocity.}
    \label{streamlines_varyingU}
\end{figure}

Figure~\ref{streamlines_varyingU} displays the velocity magnitude and streamline plots for varying Weissenberg number values. For Newtonian fluids, a perfectly steady and symmetric flow is observed, sub-figure~\ref{streamlines_varyingU}(a). The streamlines follow the pillar's body contour, and they are ordered, as expected for a creeping flow around an obstacle. A high-velocity magnitude zone is observed to be present between any two pillars present either in a horizontal or vertical line, where streamlines are merged and concentrated. Furthermore, stagnation zones with almost zero velocity magnitude are observed at the front and rear surfaces of each pillar. As fluid elasticity is introduced, noticeable changes occur even at low Weissenberg numbers. For instance, at $Wi = 1$ (sub-figure~\ref{streamlines_varyingU}(b)), the streamlines are found to be more dispersed than those seen for Newtonian fluids. The front stagnation zone is larger than the rear one. Therefore, the symmetry, particularly the vertical symmetry in the flow, is somehow lost at this Weissenberg number. Furthermore, the high-velocity magnitude formed between any two consecutive pillars is now more dispersed than that seen in Newtonian fluids.  

As the Weissenberg number further increases to 2, the flow becomes more asymmetric, as can be seen from sub-figure~\ref{streamlines_varyingU}(c). The rear stagnation zone becomes almost invisible, whereas the front stagnation zone becomes further larger. Notably, the high-velocity magnitude zone, formed between two micropillars, starts to split into two halves at this Weissenberg number, and the streamlines become further dispersed. At $Wi = 3$, a dramatic change in the flow physics is seen (sub-figure~\ref{streamlines_varyingU}(d)). First, the streamlines become highly disordered. In particular, streamlines change straight lanes and move up or down. Furthermore, the high-velocity magnitude zones have now become highly asymmetric. These zones are prominent near some micropillars, whereas they are weakened elsewhere. Therefore, some random spots of high-velocity magnitude zones form within the porous media. The stagnation zones formed near the micropillars have now also shifted to some other locations due to streamline distortion. Although the front stagnation zones are still dominant and clearly more visible than those of the rear ones. All these suggest that the flow may transit to an unsteady irregular state at this Weissenberg number. With the Weissenberg number further increased to 4, the flow irregularities become more pronounced, resulting in a highly disordered flow field, as can be seen from sub-figure~\ref{streamlines_varyingU}(e).

Interestingly, as the Weissenberg number increases further to 5 (sub-figure~\ref{streamlines_varyingU}(f)), the flow field again becomes ordered to some extent compared to that at Weissenberg numbers 3 and 4. For instance, the streamlines once again follow a straight path instead of moving up or down. In particular, curvilinear paths form between two horizontal rows of micropillars, through which most of the fluid flows, creating high-velocity zones. The rear stagnation zones again appear, although, along with the front stagnation zone, they are shifted to different locations within the micropillar compared to those observed for Newtonian and viscoelastic fluids with lower Weissenberg numbers. These suggest that the flow may again transit to a more ordered, less chaotic state than that seen at intermediate Weissenberg numbers, for instance, at $Wi = 3$ and 4.        
\begin{figure}
    \centering
    \includegraphics[width=10cm]{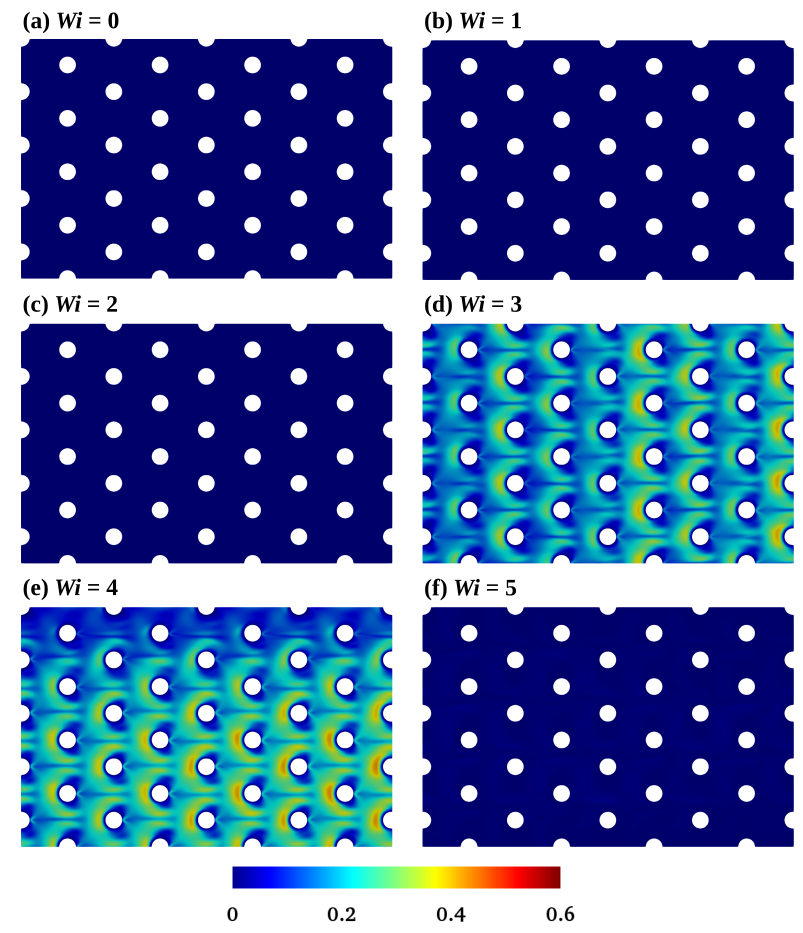}
    \caption{Contours of velocity fluctuations magnitude (normalised by $u_0$) at different Weissenberg numbers. Here, the Weissenberg number is changed by varying the mean velocity.}
    \label{Upmag_varyingU}
\end{figure}

\begin{figure}
    \centering
    \includegraphics[width=10cm]{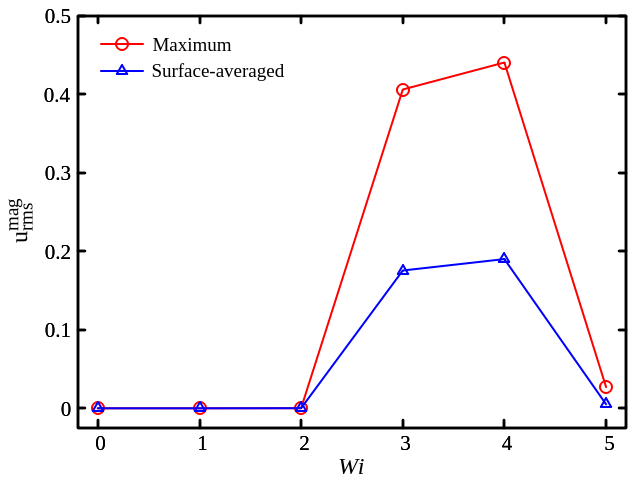}
    \caption{Maximum and surface-averaged values of $u_{\text{rms}}^{\text{mag}}$ for the domain shown in figure~\ref{Upmag_varyingU}.}
    \label{Upmag_max_sa_varyingU}
\end{figure}

To investigate and analyse this transition in the flow field with increasing $Wi$ number, we first plot the root mean square of velocity magnitude fluctuations ($u_{\text{rms}}^{\text{mag}}$) inside the porous media in figure~\ref{Upmag_varyingU}. It is defined as $u_{\text{rms}}^{\text{mag}} = \sqrt{(u_{\text{x,rms}}')^2 + (u_{\text{y,rms}}')^2}$, where $u_{\text{i,rms}}' = \sqrt{\left<(u^{\text{t}}_{\text{i}} - u^{\text{avg}}_{\text{i}})^2 \right>_t}$ with $\text{i} = \{\text{x,y} \}$, $u^{\text{t}}_{\text{i}}$ representing the velocity component at a given time `t' and $u^{\text{avg}}_{\text{i}}$ its average value. Also, the averaging of quantity $\left<. \right>$ over time is dictated by $\left<. \right>_t$. From this figure, it can be seen that for Newtonian and viscoelastic fluids with Weissenberg numbers up to 2, the value of $u_{\text{rms}}^{\text{mag}}$ is almost zero throughout the domain, suggesting that the flow field remains steady (sub-figures~\ref{Upmag_varyingU}(a)-(c)). However, as the Weissenberg number reaches 3, finite values of $u_{\text{rms}}^{\text{mag}}$ appear within the domain, particularly at the front faces of the micropillars (see sub-figure~\ref{Upmag_varyingU}(d). This indicates that the flow field starts to fluctuate at this Weissenberg number, reaching to an unsteady state. These fluctuations become more prominent as the Weissenberg number further increases to 4, as can be seen from sub-figure~\ref{Upmag_varyingU}(e). However, these fluctuations are again suppressed as the Weissenberg number further increases to 5 (sub-figure~\ref{Upmag_varyingU}(f)). This shows that a non-monotonic variation in velocity fluctuations or chaos occurs as the Weissenberg number is progressively increased. Figure~\ref{Upmag_max_sa_varyingU} further confirms this trend wherein the maximum and surface-averaged values of these velocity magnitude fluctuations are plotted against Weissenberg number. It can be seen that there is a window in Weissenberg numbers where both these parameters exhibit maximum values.

\begin{figure}
    \centering
    \includegraphics[width=14cm]{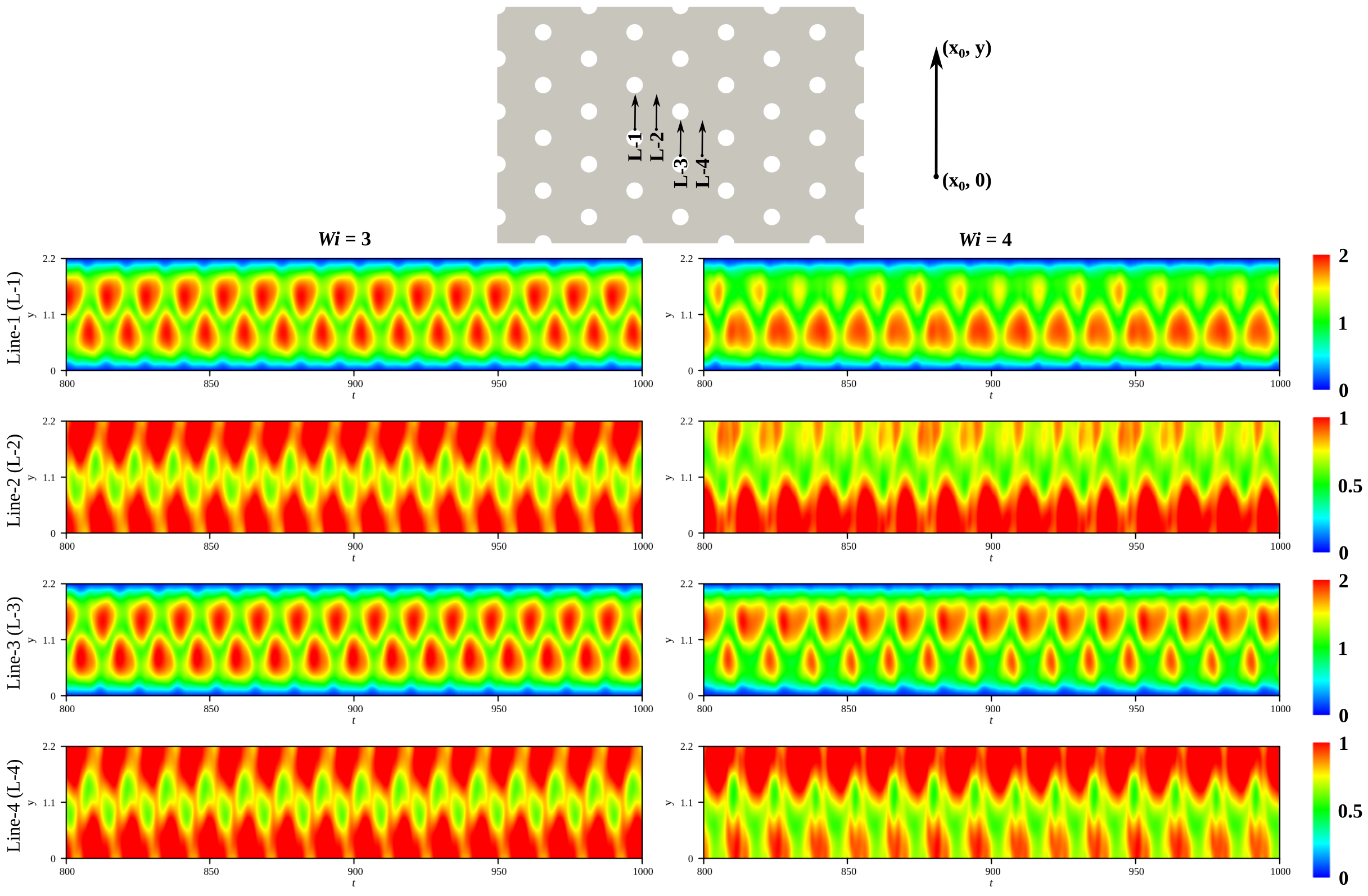}
    \caption{Kymograph of velocity magnitude (normalised by $u_0$) at four different locations (as schematically shown in the top sub-figure) for two Weissenberg number values of 3 (left half) and 4 (right half). Here, the Weissenberg number is changed by varying the mean velocity.}
    \label{kymograph_varyingU}
\end{figure}

The kymograph presented in figure~\ref{kymograph_varyingU} depicts the spatio-temporal evaluation of the velocity-magnitude field over four different vertical probe lines (as schematically shown in the same figure) present almost in the middle of the flow domain to further illustrate the transient nature of the flow field at $Wi = 3$ and 4. At $Wi=3$, the kymographs for all four probe lines exhibit highly periodic and symmetric patterns. The repeating balloon-shaped structures over the upper and lower halves of L-1 and L-3 probe lines indicate regular oscillatory motion of the flow as elastic stresses build up and relax while the fluid passes through successive micropillars. At probe lines L-2 and L-4, periodic structures become eye-shaped and more compact, indicating heterogeneity in flow fluctuations. The periodicity remains coherent over long times, suggesting that the flow disturbances are spatially organised and synchronised. As the Weissenberg number increases to $Wi=4$, the flow becomes significantly more irregular and asymmetric. The periodic structures begin to distort, merge, and fluctuate in both amplitude and spacing, especially in L-2 and L-4. This behaviour indicates stronger elastic instabilities arising from increased polymer stretching and nonlinear coupling between elastic stresses and the velocity field. The loss of temporal coherence suggests a transition toward elastic turbulence or a chaotic elastic-flow state. The differences between the probe locations are also important. Lines L-1 and L-3, which appear closer to the centre of the pore pathways, still preserve some periodicity even at $Wi=4$, although with noticeable changes. In contrast, L-2 and L-4 exhibit stronger intermittency and larger high-velocity regions, suggesting localised acceleration events and stronger elastic stress accumulation in regions influenced by pillar wakes or narrow pore throats. These regions likely experience higher extensional deformation rates, which increase polymer stretching and trigger stronger flow fluctuations. Overall, the results in figure~\ref{kymograph_varyingU} demonstrate that increasing the Weissenberg number intensifies the elastic nature of the flow, causing the system to evolve from a temporally ordered oscillatory state at $Wi=3$ to a more chaotic, spatially heterogeneous, and strongly fluctuating state at $Wi=4$.

\begin{figure}
    \centering
    \includegraphics[width=14cm]{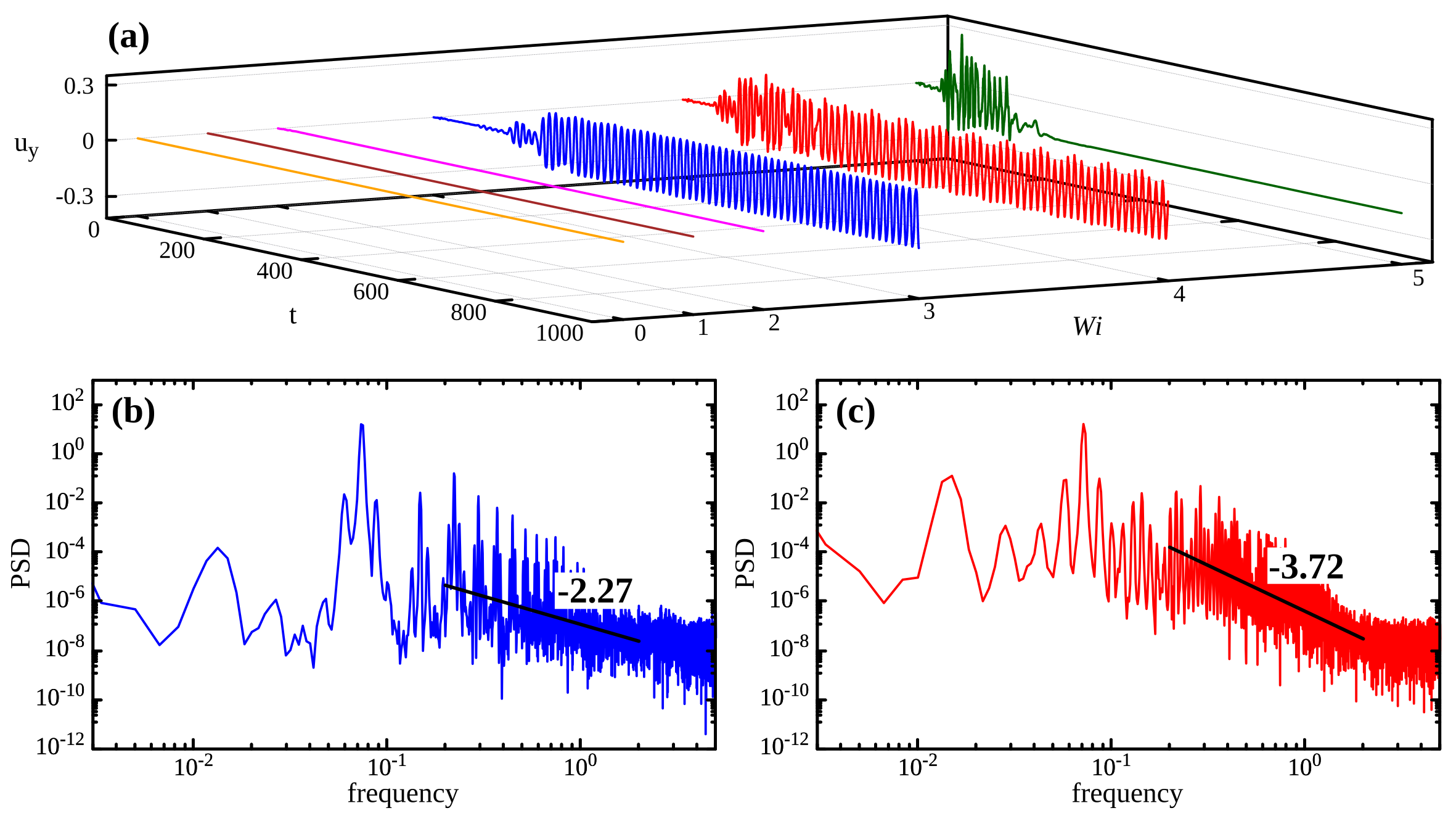}
    \caption{Time history of span-wise velocity at a probe situated at the mid-point of line-3 (L-3) in figure~\ref{kymograph_varyingU} (a). The corresponding PSD analysis is shown for $Wi = 3$ (b) and $Wi = 4$ (c). Here, the Weissenberg number is changed by varying the mean velocity.}
    \label{Uy_PSD_Probe6_varyingU}
\end{figure}

To further quantify and analyse these flow field fluctuations, their temporal evolution, along with power spectral density plots, is presented in figure~\ref{Uy_PSD_Probe6_varyingU} for a probe point located in the middle of the probe line L-3 at different Weissenberg numbers. The spanwise velocity component reaches a steady state by $Wi = 2$, indicating that the flow remains steady and symmetric under these conditions. However, at $Wi = 3$, it exhibits nearly periodic oscillations with a relatively uniform amplitude and frequency. This suggests that the flow undergoes regular transverse oscillations that remain organised in time due to cyclic stretching and relaxation of polymer molecules around the micropillar. As the Weissenberg number increases to 4, these oscillations become more irregular and complex. The amplitude changes over time, and the waveform becomes less temporally uniform, indicating increased nonlinear interaction between elastic stresses and velocity fluctuations. In contrast, at $Wi=5$, these fluctuations, which are initially present, gradually dampen over time, and the velocity component reaches almost a steady value. This suggests that a further increase in fluid elasticity causes localised flow reorganisations instead of transiting to a more chaotic state, as also evident in figure~\ref{streamlines_varyingU}.   

The corresponding PSD of velocity fluctuations at $Wi = 3$, as shown in sub-figure~\ref{Uy_PSD_Probe6_varyingU}(b), contains several sharp peaks concentrated around a dominant frequency. This indicates that the flow dynamics at this Weissenberg number is primarily dominated by a periodic oscillatory mode and its harmonics. The presence of such discrete peaks implies that only a limited number of instability modes exist at this condition. At $Wi = 4$, the PSD broadens considerably (sub-figure~\ref{Uy_PSD_Probe6_varyingU}(c)). Although some dominant peaks are still visible at this Weissenberg number, many secondary peaks also emerge, indicating stronger nonlinear interactions and energy transfer between frequencies. A plateau with a high PSD value is observed at low frequencies, corresponding to the large-scale fluctuations in the flow domain. Exponential decay $(f^{-\alpha})$ persists at high frequencies, corresponding to small-scale fluctuations in the flow domain. The decay slope is approximately $\alpha = 3.7$, thereby confirming the presence of elastic turbulence at this Weissenberg number, as also observed in prior experimental studies of flow through microporous media and other geometries~\citep{haward2021stagnation,steinberg2021elastic,sasmal2025potential,groisman2000elastic,groisman2004elastic}.

\subsection{Interlocking of birefringent strands}

The birefringent strands are highly localised regions where intense molecular stretching occurs, often near stagnation points, sharp contractions, or in the wakes of obstacles~\citep{cerf1952flow}. These strands originate from the strong coupling between flow kinematics and microstructural changes within the flow field~\citep{fuller1980flow,fuller1981flow,haward2013instabilities}. In particular, when the strain rate exceeds the relaxation rate of polymer molecules, the latter become strongly aligned and stretched in the flow direction, forming optically anisotropic "strands" visible under birefringence imaging. In polymer solutions, these strands are associated with elastic stress accumulation from the alignment and stretching of polymer molecules. Experimentally, these strands are detected using flow-induced birefringence techniques such as polarised light microscopy, crossed polarisers or birefringence retardation imaging, often combined with micro-particle image velocimetry technique~\citep{dorohoi2023review,sun2016measurements}.

\begin{figure}
    \centering
    \includegraphics[width=10cm]{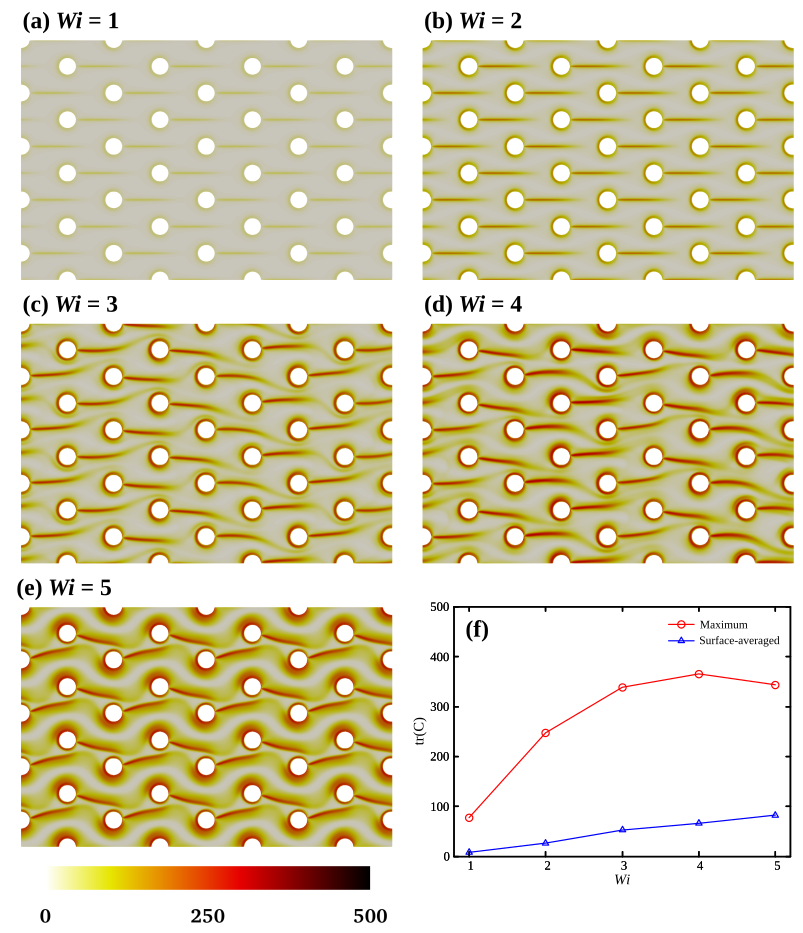}
    \caption{Contours of $tr(\bm{C})$ at different Weissenberg numbers. The sub-figure embedded at the bottom right denotes the maximum and surface-averaged values of $tr(\bm{C})$ for the domain shown in sub-figures (a)-(e). Here, the Weissenberg number is changed by varying the mean velocity.}
    \label{trC_varyingU}
\end{figure}

We, therefore, calculate the stretching of polymer molecules in the present study to locate these birefringent strands within the present flow domain. For the present FENE-CR constitutive model, the stretching of polymer molecules is the trace of polymeric conformation tensor $\bm{C}$, i.e., $tr(\bm{C})$, which is related to the polymeric stress $\bm{\tau^*_p}$ as $tr(\bm{C}) = \frac{L^2(\lambda~tr(\bm{\tau^*_p} ) + 2 \eta_p)}{\lambda~tr(\bm{\tau^*_p} ) + \eta_p L^2}$. Figure~\ref{trC_varyingU} shows the distribution of $tr(\bm{C})$ at various Weissenberg numbers ranging from 1 to 5. At $Wi = 1$, the strands start to appear downstream of the micropillars due to the high stretching of polymer molecules because of the presence of the extensional flow field in this region. The presence of such strands has also been observed experimentally in the flow of either polymer or micellar solutions past a micropillar~\citep{haward2018steady,haward2021bifurcations}. Regions of high stretching are also observed in the vicinity of the micropillar front surface, where high-shearing zones are present. As the Weissenberg number increases to 2, these strands become progressively stronger and larger due to the gradual increase in extensional flow strength downstream of the micropillars, and they remain steady and straight up to this Weissenberg number (sub-figure~\ref{trC_varyingU}(b)). As the Weissenberg number reaches 3, these strands become even stronger, but they are no longer steady in their positions; instead, they begin to oscillate in regions downstream of the micropillars, as can be seen from sub-figure~\ref{trC_varyingU}(c). The intensity of these oscillations further increases at $Wi = 4$ (sub-figure~\ref{trC_varyingU}(d)). However, with the further increase in the Weissenberg number to 5 (sub-figure~\ref{trC_varyingU}(e)), these strands initially fluctuate for some time, and after that, they no longer fluctuate; instead, they become almost stationary. In particular, the strands formed downstream of any micropillar are shifted either upward (or downward) and almost touch the micropillars in the row above (or below), and hence they become locked in their place (see the supplementary videos wherein the formation, oscillation, and locking of these strands are shown at three Weissenberg numbers, namely, 3, 4, and 5). 

\begin{figure}
    \centering
    \includegraphics[width=10cm]{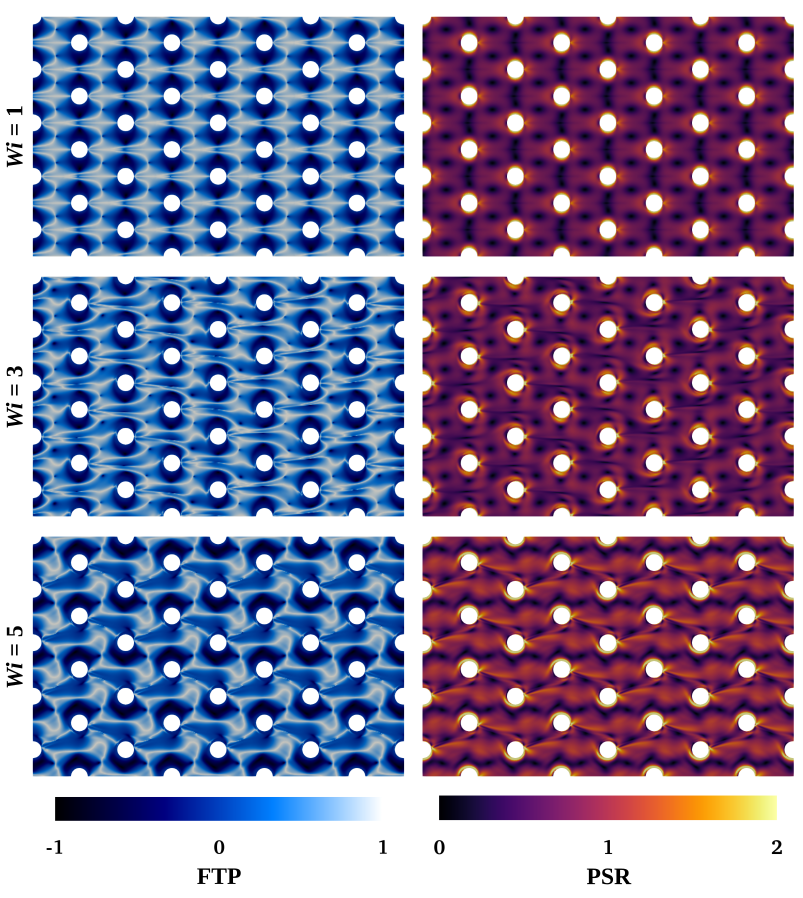}
    \caption{Flow type parameter (left half) and normalised principal strain rate (right half) at different Weissenberg numbers. Here, the Weissenberg number is changed by varying the mean velocity.}
    \label{FTP_PSR_varyingU}
\end{figure}

Sub-figure~\ref{trC_varyingU}(f) shows the maximum and surface-averaged values of $tr(\bm{C})$ as a function of the Weissenberg number quantitatively. It is observed that the maximum values gradually increase and then approach a plateau, whereas the surface-averaged values continue to increase. This increase is due to the increase in shear as well as extensional flow strengths with the Weissenberg number within the flow domain, as shown in figure~\ref{FTP_PSR_varyingU}, where the flow type parameter (left column) and principal strain rate (right column) are plotted at different Weissenberg numbers. The flow type parameter is defined as $\xi = \frac{|\mathbf{D} - |\boldsymbol{\Omega}||}{|\mathbf{D}| + |\boldsymbol{\Omega}|}$ where $|\mathbf{D}|$ and $|\boldsymbol{\Omega|}$ are the magnitudes of the strain rate and vorticity tensor, respectively~\citep{lee2007microfluidic}. The values of $\xi$ of 1, 0, and -1 represent pure extensional flows, simple shear flows, and rigid-body rotation, respectively. On the other hand, the principal strain rate (PSR) represents the extensional flow strength defined as $\frac{1}{2}\sqrt{\left(D_{xx} -D_{yy} \right)^{2} + 4D_{xy}^2}$~\citep{hamlington2008direct}. One can see that a high extensional flow field is present downstream of each micropillar, particularly in the region around the rear stagnation point, where PSR values are large. This is due to the merging of streamlines from the top and bottom sides of a micropillar, as shown in the figure~\ref{streamlines_varyingU}. As the Weissenberg number increases, the extent and intensity of these regions also increase, leading to greater stretching of polymer molecules. 

\begin{figure}
    \centering
    \includegraphics[width=10cm]{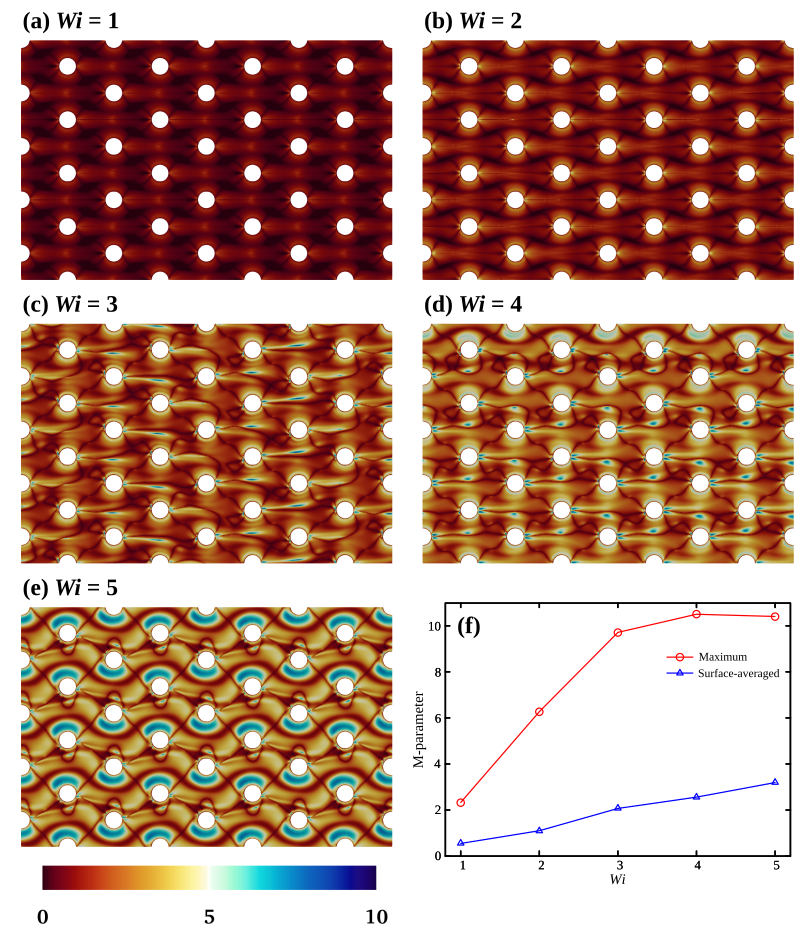}
    \caption{Variation of M-parameter at different Weissenberg numbers. The sub-figure embedded at the right bottom denotes the maximum and surface-averaged values of M-parameter for the domain shown in sub-figures (a)-(e). Here, the Weissenberg number is changed by varying the mean velocity.}
    \label{MP_varyingU}
\end{figure}

This increase in stretching of polymer molecules with the Weissenberg number results in the generation of high elastic stresses along the streamlines, which, in combination with the streamline curvature present in the porous media, lead to the onset of elastic instability within the present flow domain, which subsequently transitions to a more chaotic elastic turbulence regime. McKinley and co-workers~\citep{mckinley1996rheological,pakdel1996elastic} proposed a criterion, often termed as the Pakdel-McKinley criterion, which is used to identify the regions prone to the elastic instability in flows of viscoelastic fluids. As per their approach, they defined a dimensionless M-parameter as follows

\begin{equation} \label{eq:M-Paramter}
    M = \sqrt{\frac{\tau_{11}}{\eta_0 \dot{\gamma_1}} \frac{\lambda U}{\Re}} \geq M_{crit}
\end{equation}
Where $\tau_{11}$ is the normal stress in the flow direction along a streamline given as $\tau_{11} = \bm{t} \cdot \tau \cdot \bm{t} = \tau_{xx} t_x^2 + \tau_{yy} t_y^2 + 2\tau_{xy} t_x t_y$, where $\bm{t} = t_x \bm{e_x} + t_y \bm{e_y}$ is the tangent vector whose components can be calculated as $t_x = -\frac{\frac{\partial \psi}{\partial y}}{|\nabla \psi|}$ and $t_y = -\frac{\frac{\partial \psi}{\partial x}}{|\nabla \psi|}$. Here, $\psi$ is the stream function related to the velocity components as $u_x = -\frac{\partial \psi}{\partial y}$ and $u_y = \frac{\partial \psi}{\partial x}$, whereas $\bm{e_x}$ and $\bm{e_y}$ are the unit vectors in x and y directions, respectively. Moreover, $\dot{\gamma_1}$ in Eq.~\ref{eq:M-Paramter} signifies the characteristic value of the local deformation rate in the flow domain and $\Re$ denotes the characteristic radius of the streamline curvature given by $\kappa = {\Re^{-1}}$, where $\kappa = \frac{|\bm{u^*} \times \left(\bm{u^*} \cdot \bm{\nabla^*}\right) \bm{u^*}|}{|\bm{u^*}|^3}$. When the value of this M-parameter exceeds a critical value, elastic instability emerges in the flow domain. In the present study, this M-parameter is calculated, and its surface variation is presented in figure~\ref{MP_varyingU} at different Weissenberg numbers. It can be seen that the M-parameter becomes noticeable at the front and rear stagnation points of the micropillars at $Wi = 2$. With an increase in the Weissenberg number, it becomes pronounced at the rear stagnation points and in the regions between two micropillars; for instance, see the results at $Wi = 3$ and 4 in sub-figures~\ref{MP_varyingU}(d) and (e), respectively. This is mainly due to the presence of high streamline curvature in both regions. Therefore, these two regions become highly susceptible to generating elastic instability and elastic turbulence. When the Weissenberg number increases further to 5, the region with high M-parameter also increases; however, it now shifts toward the side of the micropillar, which is arranged in a zig-zag manner (see sub-figure~\ref{MP_varyingU}). This is again caused by the high streamline curvature at this Weissenberg number due to the formation of a curvilinear flow path, as seen in sub-figure~\ref{streamlines_varyingU}. Although the region with the highest M-parameter value further increases at this Weissenberg number, the flow field is found to be stable at this Weissenberg number due to the interlocking of birefringent strands. Interestingly, this is a counterintuitive situation where the flow field becomes stable even with a spot of high M-parameter value. Sub-figure~\ref{MP_varyingU}(f) presents the corresponding maximum and surface-averaged values of the M-parameter. One can see that the maximum value saturates at higher Weissenberg numbers; however, the surface-averaged value increases progressively with the Weissenberg number due to an increase in the extent of the region with high M-parameter.

\subsection{Pressure drop variation}

In this subsection, we aim to calculate the variation in pressure drop across the porous media. In doing so, multiplying both sides of the momentum equation (equation~\ref{eq:momentum_dimensional}) by the velocity vector $\bm{u^*}$, we get

\begin{equation}\label{eq:momentum_multiply_with_velocity}
\bm{u^*} \cdot \rho \left( \frac{\partial \bm{u^*}}{\partial t^*} + \bm{u^*} \cdot \bm{\nabla^*} \bm{u^*} \right) = - \bm{u^*} \cdot \bm{\nabla^*} p^* + \bm{u^*} \cdot \bm{\nabla^*} \cdot \bm{\tau^*}
\end{equation}

\noindent
Now using the identities $
\bm{u^*} \cdot (\bm{u^*} \cdot \bm{\nabla^*} \bm{u^*}) = \bm{\nabla^*} \cdot \left( \frac{1}{2} (u^*)^2 \bm{u^*} \right)$ where $(u^*)^2 = \bm{u^*} \cdot \bm{u^*}$ and $
\bm{u^*} \cdot \frac{\partial \bm{u^*}}{\partial t^*} = \frac{\partial}{\partial t^*}\left(\frac{1}{2}(u^*)^2\right)$, we get the following form of the equation~\ref{eq:momentum_multiply_with_velocity}

\begin{equation}
\frac{\partial}{\partial t^*}\left(\frac{1}{2}\rho (u^*)^2\right) + \bm{\nabla^*} \cdot \left(\frac{1}{2}\rho (u^*)^2 \bm{u^*}\right) = - \bm{u^*} \cdot \bm{\nabla^*} p^* + \bm{u^*} \cdot \bm{\nabla^*} \cdot \bm{\tau^*}
\end{equation}

\noindent
Now applying vector identities

\begin{equation}
  - \bm{u^*} \cdot \bm{\nabla^*} p^* = - \bm{\nabla^*} \cdot (p^* \bm{u^*}) + p^* \, \bm{\nabla^*} \cdot \bm{u^*}
\end{equation}

\noindent
Since the present flow is incompressible, i.e., $\bm{\nabla^*} \cdot \bm{u^*} = 0$, we thus get

\begin{equation}
    - \bm{u^*} \cdot \bm{\nabla^*} p^* = - \bm{\nabla^*} \cdot (p^* \bm{u^*})
\end{equation}

\noindent
Also, by using the following identity

\begin{equation}
    \bm{u^*} \cdot \bm{\nabla^*} \cdot \bm{\tau^*} = \bm{\nabla^*} \cdot (\bm{\tau^*} \cdot \bm{u^*}) - \bm{\tau^*} : \bm{\nabla^*} \bm{u^*}
\end{equation}

\noindent
We ultimately get the following equation

\begin{equation}\label{eq:momentum_multiply_with_velocity_simplified}
    \frac{\partial}{\partial t^*} \left(\frac{1}{2} \rho (u^*)^2 \right) + \bm{\nabla^*} \cdot \left(\frac{1}{2} \rho (u^*)^2 \bm{u^*} \right) + \bm{\nabla^*} \cdot \left(p^* \bm{u^*} \right) - \bm{\nabla^*} \cdot \left(\bm{\tau^*} \cdot \bm{u^*} \right) = -\bm{\tau^*} : \bm{\nabla^*} \bm{u^*}
\end{equation}
The first term on the left-hand side of equation~\ref{eq:momentum_multiply_with_velocity_simplified} represents the rate of change of kinetic energy (KE); the second, third, and fourth terms represent the energy fluxes due to kinetic energy, pressure, and stress, respectively. The right-hand side term represents the dissipation, which has two contributions, namely, solvent contribution $(\epsilon_s)$, and polymeric contribution $(\epsilon_p)$. The first two terms are negligible in a low-Reynolds-number flow compared to the other terms in the equation above. This can be proved from a simple scaling argument as follows: the KE scales as $\sim \rho \frac{u^3_0}{D_0}$ whereas the viscous dissipation scales as $\sim \eta_0 \frac{u^2_0}{D^2_0}$. If we take the ratio of these two contributions, we simply get $\frac{\rho \frac{u^3_0}{D_0}}{\eta_0 \frac{u^2_0}{D^2_0}} = \frac{\rho u_0 D_0}{\eta_0} = Re$, which is nothing but the Reynolds number, which is $<<1$ in the present case. Therefore, the above equation is simplified as follows

\begin{equation}\label{eq:momentum_multiply_with_velocity_simplified_further}
    \bm{\nabla^*} \cdot \left(p^* \bm{u^*} \right) - \bm{\nabla^*} \cdot \left(\bm{\tau^*} \cdot \bm{u^*} \right) = - \bm{\tau^*} : \bm{\nabla^*} \bm{u^*}
\end{equation}
Furthermore, the contribution of the second term on the left-hand side of equation~\ref{eq:momentum_multiply_with_velocity_simplified_further} is negligible compared to the first term. This can be again proved from a simple scaling argument: for porous media flows, the pressure work per unit area scales as $p^* u_0 \sim \frac{\eta_0 u^2_0 L}{D^2_0}$, where $L$ is the length of the porous media, whereas the viscous traction power scales as $\bm{\tau^*} u_0 \sim \frac{\eta_0 u^2_0}{D_0}$. If we take the ratio of two, we get $\frac{L}{D_0}$, where $L >> D_0$, and therefore, we can neglect the last term in the left-hand side of equation~\ref{eq:momentum_multiply_with_velocity_simplified_further} compared to the first term. Finally, the energy equation becomes

\begin{equation}
       -\bm{\nabla^*} \cdot \left(p^* \bm{u^*} \right) = \bm{\tau^*} : \bm{\nabla^*}\bm{u^*}
\end{equation}
Integrating the above equation over a control volume and applying the divergence theorem, we can get

\begin{equation}
    \Delta p^* Q = \int_V \bm{\tau^*} : \bm{\nabla} \bm{u^*}\, dV
\end{equation}
Where $\Delta p^*$ is the pressure drop and $Q$ is the volumetric flow rate. Substituting $\bm{\tau^*} = \bm{\tau^*_s} + \bm{\tau^*_p} = 2 \eta_s \dot{\bm{\gamma^*}} + \bm{\tau^*_p}$, we get

\begin{equation}
     \Delta p^* Q = \int_V \left[(2 \eta_s (\dot{\bm{\gamma^*}} : \dot{\bm{\gamma^*}})) + (\bm{\tau^*_p} : \dot{\bm{\gamma^*}})\right]\, dV
\end{equation}
For a viscoelastic fluid, $\bm{\tau^*_p}: \dot{\bm{\gamma^*}} = \frac{D W^*}{D t^*} + \epsilon_p$, where the first term on the right-hand represents elastic stored energy and the second term represents the polymeric dissipated energy~\citep{snoeijer2020relationship}. For a statistically stationary flow, $\left<\frac{D W^*}{D t^*}\right> = 0$, and therefore, we get

\begin{equation}
    \Delta p^* Q = \int_V \left[ 2 \eta_s (\dot{\bm{\gamma^*}} : \dot{\bm{\gamma^*}}) + \epsilon_p \right]\, dV
\end{equation}
For the FENE-CR model, the elastic dissipation energy is $\epsilon_p = \frac{\eta_P}{\lambda}f(I_1) \left[ I_1 f(I_1) -3 \right]$, where $I_1$ is the trace of the conformation tensor, i.e., $I_1 = tr(\mathbf{C})$ and $f(I_1) = \frac{L^2 - 3}{L^2 - I_1}$~\citep{snoeijer2020relationship}. After performing some rearrangements, we can write the above equation as 

\begin{equation}\label{eq:pressureDrop}
    \frac{\Delta p^*}{L}  = \frac{\left<\text{VD} \right>_V + \left<\text{PD} \right>_V}{u_0}
\end{equation}
Where $\text{VD}$ is the viscous dissipation, $\text{PD}$ is the polymeric dissipation, and $\left<. \right>_V$ denotes the volume-averaged quantity. These volume-averaged quantities will have two components, namely, mean and fluctuating component, for statistically stationary flow, i.e., $\left<\text{VD} \right>_V = \left<\overline{\text{VD}} \right>_V + \left<\text{VD}^{'} \right>_V$ and $\left<\text{PD} \right>_V = \left<\overline{\text{PD}} \right>_V + \left<\text{PD}^{'} \right>_V$.    

\begin{figure}
    \centering
    \includegraphics[width=10cm]{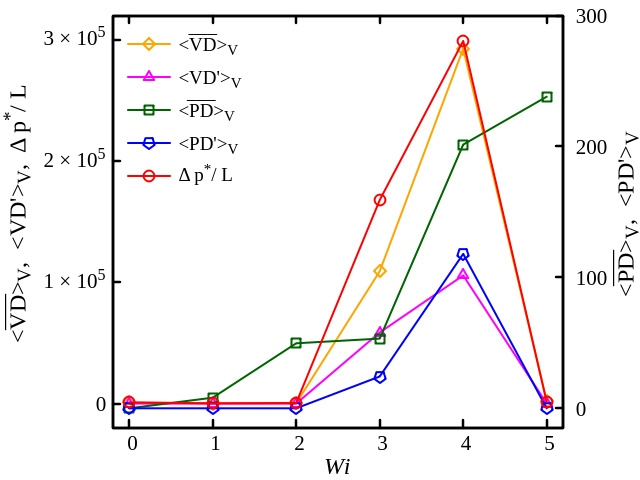}
    \caption{Variation of various parameters involved in Eq.~\ref{eq:pressureDrop} with different Weissenberg number values. Here, the Weissenberg number is changed by varying the mean velocity.}
    \label{deltaP_varyingU}
\end{figure}

Figure~\ref{deltaP_varyingU} depicts the variation in pressure drop with the Weissenberg number, along with four different contributions. It can be seen that the pressure drop shows a non-monotonic trend with the Weissenberg number, i.e., it first increases, reaches a maximum, and then decreases as the Weissenberg number increases gradually, as also observed by Browne and Datta~\citep{browne2021elastic} in their experiments through a three-dimensional porous medium composed of randomly packed spheres. The four corresponding contributions to this pressure drop exhibit different trends with the Weissenberg number. The mean viscous contribution first decreases, then increases to a maximum, and then decreases again with the Weissenberg number, similar to the pressure drop. On the other hand, the mean polymeric contribution increases as the Weissenberg number increases, due to an increase in the stretching of polymer molecules, as also evident in figure~\ref{trC_varyingU}. In contrast, the fluctuating contributions show a non-monotonic trend with the Weissenberg number. They are small up to $Wi = 2$, and then become very large at $Wi = 3$ and increase further at $Wi = 4$. However, at $Wi = 5$, their contributions again become small due to the suppression of chaotic flow dynamics at this Weissenberg number because of the interlocking of birefringent strands as discussed in the preceding subsection. Therefore, the non-monotonicity in the pressure drop variation with the Weissenberg number appears due to the non-monotonic variation in flow field fluctuations.

\begin{figure}
    \centering
    \includegraphics[width=10cm]{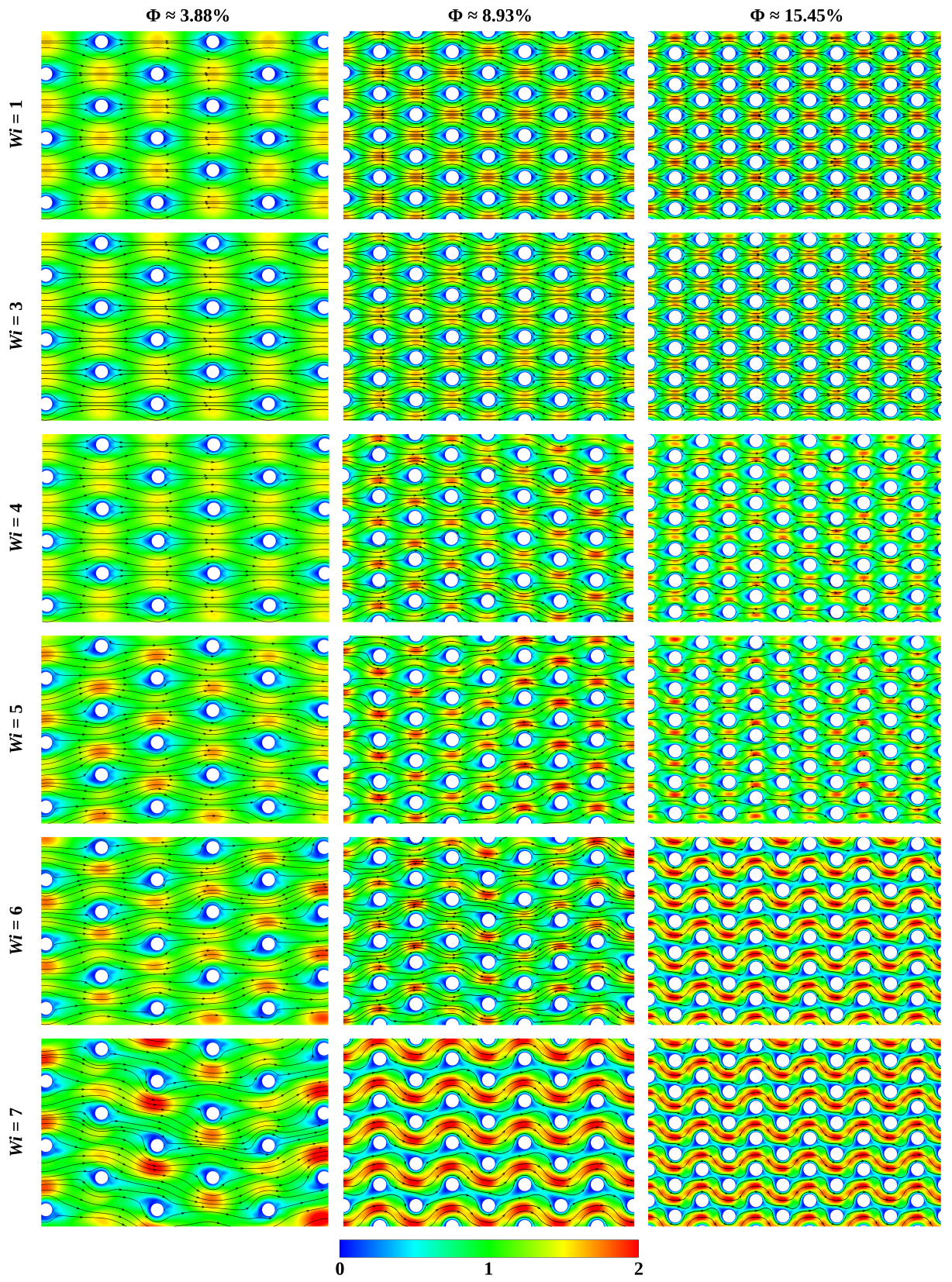}
    \caption{Streamlines and velocity magnitude (normalised by $u_0$) plots at different Weissenberg numbers and solid volume fractions of the porous media.}
    \label{streamlines}
\end{figure}

\begin{figure}
    \centering
    \includegraphics[width=10cm]{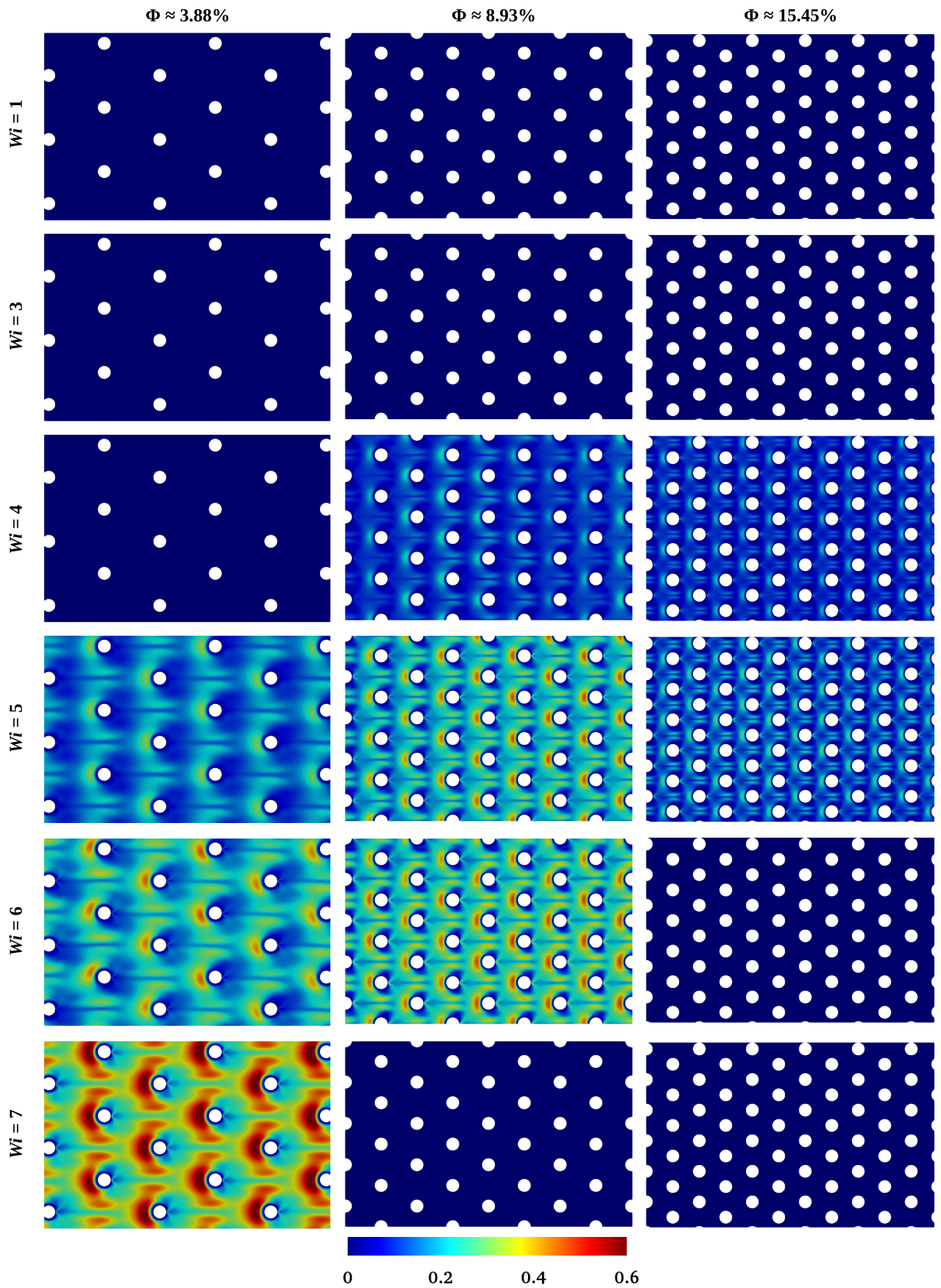}
    \caption{Contours of velocity fluctuations magnitude (normalised by $u_0$) at different Weissenberg numbers and solid volume fractions of the porous media.}
    \label{Upmag}
\end{figure}

\begin{figure}
    \centering
    \includegraphics[width=10cm]{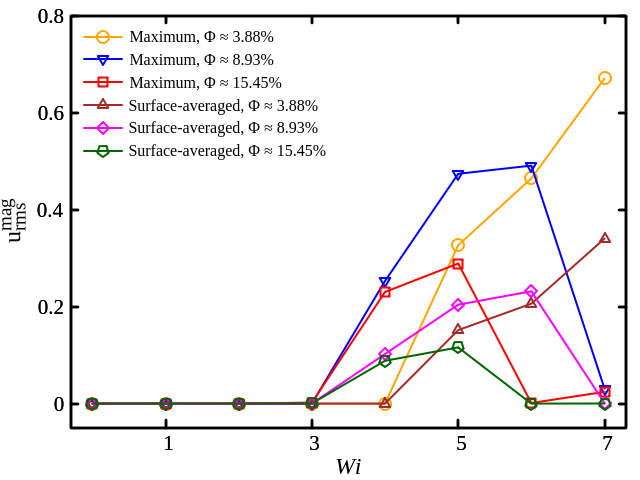}
    \caption{Maximum and surface-averaged values of $u_{\text{rms}}^{\text{mag}}$ for different solid volume fractions of the porous media within the domain shown in figure~\ref{Upmag}.}
    \label{Upmag_max_sa}
\end{figure}

\begin{figure}
    \centering
    \includegraphics[width=10cm]{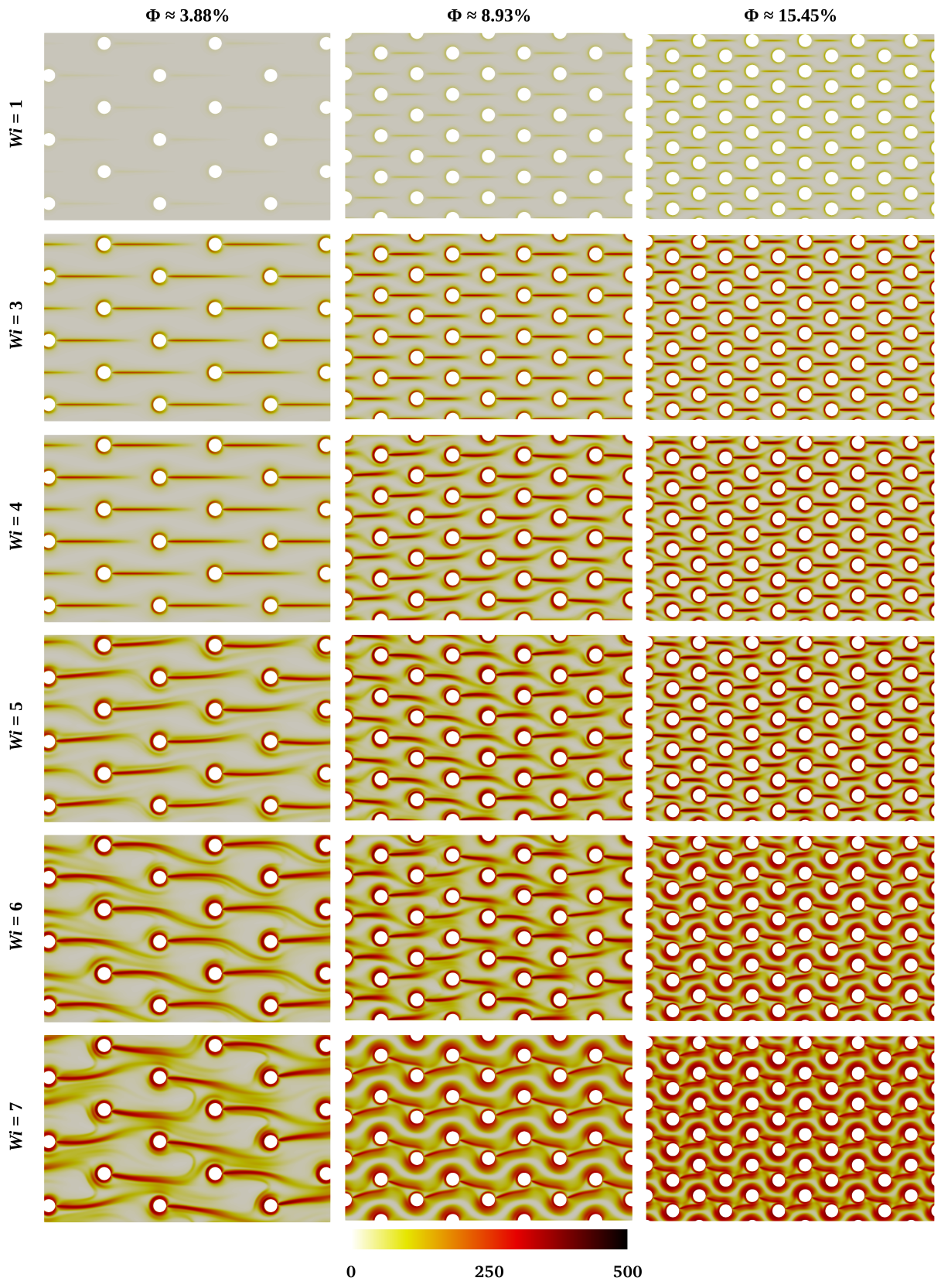}
    \caption{Contours of $tr(\bm{C})$ at different Weissenberg numbers and solid volume fractions of the porous media.}
    \label{tr(C)}
\end{figure}

\begin{figure}
    \centering
    \includegraphics[width=10cm]{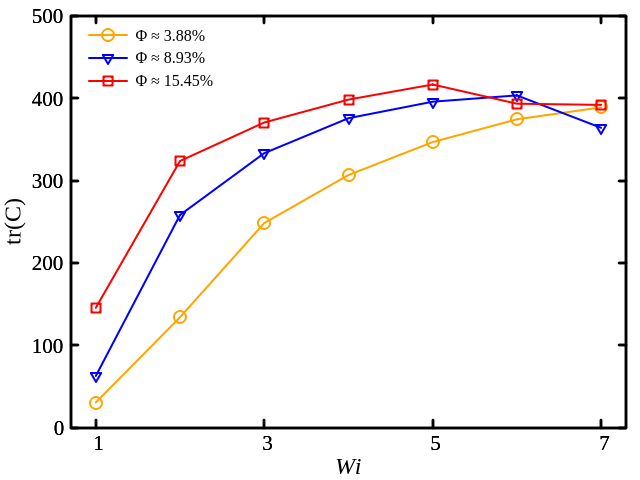}
    \caption{Maximum values of $tr(\bm{C})$ for different solid volume fractions of the porous media within the domain shown in figure~\ref{tr(C)}.}
    \label{trC_max}
\end{figure}

\subsection{Effect of solid volume fraction}
We further run simulations for different solid volume fraction percentages (denoted by $\Phi$ and defined as $V_S/V_T \times 100$, where $V_T$ and $V_S$ are the total volume of the domain and total solid volume in the domain, respectively) within the porous media to examine how they influence the associated non-monotonicity in the flow behaviour. Note that this set of simulations is run with the non-dimensional elastic modulus $G = \left(\frac{1-\beta}{Wi}\right)$ of the viscoelastic fluid held constant at 0.05. For example, the Weissenberg number is increased by increasing the relaxation time $\lambda$ and decreasing the viscosity ratio $\beta$ such that the value of $G$ remains constant. The value of $u_0$ is held constant in this case. Figure~\ref{streamlines} represents the streamlines and velocity magnitude plots at three different solid volume fractions within the porous media and at different Weissenberg numbers ranging from 1 to 7. Naturally, an increase in the solid volume fraction reduces the flow area, resulting in a larger velocity-magnitude zone in the open area, particularly between two micropillars. Likewise, as seen earlier, for Newtonian and viscoelastic fluids with low Weissenberg numbers, for instance, at $Wi = 1$, the flow field remains smooth, symmetric, and aligned with the mean flow direction for all solid volume fractions. Furthermore, with the gradual increase in the Weissenberg number, the flow field progressively loses its symmetry for all solid volume fractions. The splitting of the high-velocity magnitude zone, situated between two micropillars, into two parts begins again for all solid volume fractions of the porous media. However, as the Weissenberg number further increases, a noticeable change in the flow dynamics is observed, depending on the solid volume fraction of the porous medium. At the lowest $\Phi$, the flow field still remains in some organised state where high-velocity magnitude zones look the same, and streamlines follow the mean flow direction at $Wi = 4$. At the same Weissenberg number, in contrast, at the highest $\Phi$, a surface variation in the intensity of the high-velocity magnitude zone is seen, and streamlines also become distorted, particularly deviating from their mean flow direction and moving either upwards or downwards. This suggests that the flow field may transit to an unsteady state at this solid volume fraction of the porous medium. As the Weissenberg number further increases to 5, such a transition is even seen for the lowest $\Phi$. Therefore, the transition from steady to unsteady flow due to elastic instability is accelerated with increasing solid volume fraction in the porous medium.  

The chaotic nature of the flow field further increases as the Weissenberg number increases to 6 for $\Phi = 3.88\%$ and $8.93\%$, as evident from the irregularities in the distribution of the intensity of the high-velocity magnitude zone as well as their shapes. In contrast, for $\Phi = 15.45\%$, the flow field starts to reorganise as the formation of curvilinear flow paths is seen at this solid volume fraction, indicating that the flow may again be transitioning to the steady state. This reorganisation of flow is also seen at $\Phi = 8.93\%$ when the Weissenberg number further increases to 7, whereas the corresponding flow becomes more irregular and chaotic at $\Phi = 3.88\%$. To quantify the unsteadiness or chaos in the flow, we present the surface distribution of the root-mean-square velocity magnitude fluctuation over the domain in figure~\ref{Upmag}. The intensity of flow field fluctuations, primarily near the front surface of each micropillar, gradually increases with the Weissenberg number at the lowest solid volume fraction considered in this study (see the first column in figure~\ref{Upmag}). In contrast, the fluctuations exhibit a non-monotonic trend for $\Phi = 8.93\%$ and $15.45\%$. This is quantitatively shown in figure~\ref{Upmag_max_sa}, wherein the maximum and surface-averaged values of $u^{\text{mag}}_{\text{rms}}$ are plotted against the Weissenberg number at different solid volume fractions. It can be seen that the onset of fluctuation is delayed to a higher Weissenberg number as the solid volume fraction decreases. For the lowest solid volume fraction, both the maximum and surface-averaged values increase with Weissenberg number once the critical Weissenberg number is exceeded. In contrast, a window in the Weissenberg number exists for the other two solid volume fractions, in which fluctuations in both maximum and surface-averaged values remain finite and non-monotonic. However, the window size decreases as the solid volume fraction increases, suggesting that the flow field's chaotic nature decreases.    

The corresponding stretching of polymer molecules is shown in figure~\ref{tr(C)} at different Weissenberg numbers and solid volume fractions of the porous media. The birefringent strands formed downstream of the micropillars remain straight in the mean flow direction and steady up to $Wi = 3$ for all solid volume fractions. However, as the Weissenberg number increases to 4, they remain in the same state for $\Phi = 3.88\%$, but start to oscillate and deviate from their straight positions for the other two higher solid volume fractions. With further increase in the Weissenberg number, the interlocking of birefringent strands happens for $\Phi = 8.93\%$ and $15.45\%$, whereas it does not happen for $\Phi = 3.88\%$. The maximum values of polymer stretching are shown in the figure~\ref{trC_max}. As the solid volume fraction increases, the stretching of polymer molecules also increases at a fixed Weissenberg number due to the increase in extensional flow strength. They also increase gradually with the Weissenberg number; however, they ultimately reach a plateau value that is nearly the same at high Weissenberg numbers, irrespective of $\Phi$. Despite achieving almost the same maximum stretching and longer birefringent strand lengths, interlocking does not occur at the lowest solid volume fraction, which could stabilise the flow at high Weissenberg numbers. This is because the birefringent strands have a larger volume for their movements in this case, which decreases gradually as the solid volume fraction increases. This further confirms our hypothesis that the interlocking of birefringent strands is the controlling factor in modulating chaos and, subsequently, the pressure drop across viscoelastic fluid flow through a porous medium. 


\section{Conclusions}
This study presents a numerical investigation of viscoelastic fluid flows through a model microporous medium consisting of micropillars arranged in a staggered pattern over a wide range of Weissenberg numbers. A transition in the flow field occurs in this flow geometry as the Weissenberg number increases. At low Weissenberg numbers, the flow field exhibits a steady, ordered, and symmetric pattern. However, as the Weissenberg number increases, it becomes progressively asymmetric due to elastic instabilities governed by normal elastic stresses, ultimately transitioning to a chaotic elastic turbulent regime. However, with further increase in the Weissenberg number, we find that the flow field reorganises into an ordered and almost steady state, thereby suppressing chaos at high Weissenberg numbers. Therefore, a non-monotonic variation in the flow-field fluctuations is observed with the increasing value of the Weissenberg number. The same trend is also observed in the corresponding pressure drop variation across the porous media, which arises from two contributions: a mean contribution from statistically stationary flow quantities and a fluctuating contribution. While the mean contribution gradually increases with the Weissenberg number, the fluctuating contribution exhibits non-monotonic variation, ultimately leading to a non-monotonic variation of the total pressure drop. Such flow behaviours are also seen in recent experiments, both in ordered two-dimensional and random three-dimensional porous media~\citep{haward2021stagnation,browne2021elastic}. 

We propose that this non-monotonic variation, both in chaotic flow behaviour and in pressure drop, is governed by the formation and evaluation of birefringent strands of high elastic stress, arising from the stretching and alignment of polymer molecules in the flow domain. These strands develop gradually as the Weissenberg number increases and begin to fluctuate within the space between the obstacles once the Weissenberg number exceeds a critical value, thereby causing the flow field to transition to a chaotic elastic-turbulent regime. However, as the Weissenberg number increases further, although these strands get stronger and larger, they become interlocked with neighbouring obstacles, thereby suppressing the chaotic flow dynamics at high Weissenberg numbers. Due to this interlocking of strands, the fluctuating contribution to the pressure drop also decreases at higher Weissenberg numbers, resulting in the same non-monotonic variation. Furthermore, we show that the transition in the flow field from steady to unsteady chaotic strongly depends on the solid volume fraction of the porous media. In particular, as the solid volume fraction increases, the onset of instabilities is accelerated at lower Weissenberg numbers; however, the overall fluctuation intensity decreases as the interlocking of birefringent strands becomes accentuated. All in all, the findings of the present study can guide the control of the chaotic flow behaviour of viscoelastic fluids in porous media, which has the potential to be harnessed to increase chemical reaction rates or displace other fluids. 

\section*{Acknowledgments}
We acknowledge the National Supercomputing Mission (NSM) for providing computing resources of ‘PARAM Smriti’ at NABI, Mohali (accessed by CS) and ‘PARAM Himalaya’ at IIT Mandi (accessed by AC), which are implemented by C-DAC and supported by the Ministry of Electronics and Information Technology (MeitY) and Department of Science and Technology (DST), Government of India. The HPC facility of IIT Ropar is also commended. CS acknowledges financial support from the Anusandhan National Research Foundation (ANRF), Government of India, through a core research grant (CRG/2023/001908), and AC thanks the Ministry of Education, Government of India, for financial support from the PMRF (Cycle-9).

\section*{Supplementary Material}
Supplementary video on the evaluation dynamics of birefringent strands at three Weissenberg numbers, namely, 3, 4, and 5.   

\section*{Declaration of Interests}
The authors report no conflict of interest.



\bibliographystyle{jfm}
\bibliography{jfm}
\end{document}